\documentclass[journal]{IEEEtran}
\usepackage{cite}
\usepackage{amsmath,amssymb,amsfonts,amsthm}

\usepackage[utf8]{inputenc} 
\usepackage[T1]{fontenc}

\usepackage{graphicx}
\usepackage{textcomp}
\usepackage{xcolor,soul}
\usepackage{stmaryrd}
\usepackage{algpseudocode}
\usepackage{amsmath}
\usepackage{mathrsfs}
\usepackage{bm}
\usepackage[abs]{overpic}
\usepackage[nocomma]{optidef}

\usepackage{physics}
\usepackage{tikz}
\usepackage{mathdots}
\usepackage{yhmath}
\usepackage{cancel}
\usepackage{siunitx}
\usepackage{array}
\usepackage{multirow}
\usepackage{amssymb}
\usepackage{gensymb}
\usepackage{tabularx}
\usepackage{extarrows}
\usepackage{booktabs}
\usepackage[ruled,linesnumbered]{algorithm2e}
\makeatletter
\newcommand{\mathleft}{\@fleqntrue\@mathmargin0pt}
\newcommand{\mathcenter}{\@fleqnfalse}
\makeatother

\usetikzlibrary{fadings}
\usetikzlibrary{patterns}
\usetikzlibrary{shadows.blur}
\usetikzlibrary{shapes}

\theoremstyle{plain}
\newtheorem*{theorem*}{Theorem}

\newtheorem{remark}{Remark}
\newtheorem*{eg}{Example}

\usepackage{xcolor,varwidth}

\usepackage[scaled=.75]{beramono}
\newcommand{\bpara}[1]		{\medskip \noindent {\bf #1}}

\renewcommand\geq\geqslant
\renewcommand\leq\leqslant

\newcommand\fig[1]			{Fig.~\ref{#1}}

\def\eg					    {\emph{e.g.,}~}
\def\ie					    {i.e., ~}

\renewcommand{\Re}[1]{\mathfrak{R}\left\{#1\right\}}

\usepackage{xspace}

\DeclareDocumentCommand{\sd}{o}  
{{\underline{\ast}\IfValueT{#1}{_{#1}}}}

\DeclareDocumentCommand{\xmod}{m o o o}  
{%
\IfNoValueTF{#4}
{{#1}\IfValueT{#2}{_{m,\mathsf{#2}}}\IfValueT{#3}{#3}}
{{#1}\IfValueT{#2}{_{m,\mathsf{#2}}^{\mathsf{#4}}}\IfValueT{#3}{#3}}
}

\usepackage[inline]{enumitem}
\DeclareMathAlphabet{\mathsfit}{T1}{\sfdefault}{\mddefault}{\sldefault}
\SetMathAlphabet{\mathsfit}{bold}{T1}{\sfdefault}{\bfdefault}{\sldefault}
\usepackage[switch]{lineno}
\usepackage{siunitx}

\def\BibTeX{{\mathrm B\kern-.05em{\sc i\kern-.025em b}\kern-.08em
    T\kern-.1667em\lower.7ex\hbox{E}\kern-.125emX}}
    
\usepackage{hyperref}
\usepackage{etoolbox}

\begin{document}

\title{
Enabling Full-Duplex LEO Satellite Systems with Non-Reciprocal BD-RIS-Assisted Beamforming 
}


\author{Ziang Liu,
        Wonjae Shin,~\IEEEmembership{Senior Member,~IEEE}
        and~Bruno Clerckx,~\IEEEmembership{Fellow,~IEEE}

\thanks{Z. Liu, and B. Clerckx are with the Communications \& Signal Processing (CSP) Group at the Department of Electrical and Electronic Engineering., Imperial College London, SW7 2AZ, UK. (e-mails:\{ziang.liu20, b.clerckx\}@imperial.ac.uk). 

W. Shin is with the School of Electrical Engineering, Korea University, Seoul 02841, South Korea (e-mail:
wjshin@korea.ac.kr).}

\thanks{Bruno Clerckx is also with Kyung Hee University, Seoul, Korea. Corresponding author: Bruno Clerckx.}
}


\maketitle

\begin{abstract}
    Low Earth orbit (LEO) satellites are a promising technology for providing low-latency, high-data-rate, and wide-coverage communication services. However, with growing demand for data transmission, future non-terrestrial networks (NTNs) require high spectral efficiency especially with low-gain antennas at the ground devices. This motivates the adoption of in-band full-duplex (FD) systems. In addition, the potential imbalance between downlink (DL) and uplink (UL) transmissions necessitates flexibility in resource allocation. To overcome these challenges, we propose an FD LEO satellite system, where the non-reciprocal beyond-diagonal reconfigurable intelligent surfaces (NR-BD-RIS) and multiple transmit and receive antennas are attached to the LEO satellite. NR-BD-RIS reflects the DL and UL signals by passive beamforming. By incorporating non-reciprocal components into the impedance network of RIS, the NR-BD-RIS breaks channel reciprocity, facilitating simultaneous support for multiple beam directions. To cover a wide coverage, we propose a time-sharing scheduling framework in which the NR-BD-RIS simultaneously serves multiple DL and multiple UL ground devices within each time slot. An optimization problem is defined to maximize the weighted sum-rate over the entire scheduling period. Numerical results demonstrate that the proposed NR-BD-RIS significantly performs better than both conventional BD-RIS and diagonal RIS (D-RIS) with respect to DL and UL sum-rate performance under both single-user (SU) and multiple-user (MU) cases. Additionally, NR-BD-RIS requires less frequent reconfiguration compared to the other two types of RIS, making it more practical for implementation.
\end{abstract}


\section{Introduction}
Low Earth orbit (LEO) satellite system, operating at altitudes of $500$–$2000$ km, offer low latency, high data rates, and wide coverage, making it a promising solution for ubiquitous, high-capacity global connectivity in current and future wireless networks \cite{toka2024ris}. {However, future non-terrestrial networks (NTNs) demand efficient spectrum use especially with low-gain antennas at ground devices. This motivates in-band full-duplex (FD) operation \cite{lagunas2025full}, which can potentially double spectral efficiency by enabling simultaneous transmission and reception. In addition, the potential imbalance between downlink (DL) and uplink (UL) transmissions requires flexibility in resource allocation. These factors motivate the need for FD LEO satellite systems that provide low cost, high spectrum usage and flexibility to tackle the DL and UL imbalance.}

Reconfigurable intelligent surfaces (RIS) have obtained extensive attention in both academia and industry due to their capability to intelligently manipulate the wireless propagation environment, therefore enabling high energy and spectrum efficiency \cite{huang2019reconfigurable}. Furthermore, RIS is a low-cost planar surface consisting of passive, reconfigurable scattering elements, making its integration into LEO satellites a promising approach to satisfy the strict requirements of FD LEO satellite systems \cite{toka2024ris}. Several studies have examined the utilization of RIS for improving LEO satellite communication performance. For instance, \cite{toka2024ris} proposes a RIS-assisted LEO satellite system to achieve energy-efficient transmission. Additionally, a delay-adjustable RIS (DA-RIS) is introduced in \cite{sekimori2024frequency} to realize a frequency prism, addressing size, weight, and power (SWaP) constraints while improving beam flexibility.

From a technical perspective, we can use an array of scattering elements connected to reconfigurable impedance network to model RIS. In the diagonal RIS (D-RIS), each port of the impedance network is connected to ground through a single impedance component. This architecture, also referred to as single-connected RIS \cite{shen_modeling_2022}, leads to a diagonal scattering matrix. {Consequently, the D-RIS can only adjust the phase of the incident wave at each port, thereby limiting its passive beamforming capabilities.}
To address the limitations of controlling only the phase of the incident wave, \cite{shen_modeling_2022, li_reconfigurable_2024, li2025tutorial} introduce beyond-diagonal (BD)-RIS. Different from D-RIS, the impedance network ports of BD-RIS are interconnected, allowing the output wave at one port to depend not only on the wave impinging on that port but also on waves from neighboring ports. These interconnections lead to a non-diagonal scattering matrix. BD-RIS has demonstrated significant improvements in spectrum efficiency for wireless communications \cite{shen_modeling_2022, li_reconfigurable_2024}, and improved performance in sensing and localization systems \cite{liu2024enhancing}. However, the interconnection of all ports and the required impedance components increase circuit complexity. 
To achieve a trade-off between hardware complexity and system performance, prior work \cite{shen_modeling_2022} has introduced the group-connected (GC)-BD-RIS, where ports are interconnected within groups, leading to a block-diagonal scattering matrix. 
Further advancements include tree-connected and forest-connected BD-RIS structures proposed in \cite{nerini2024beyond}, which demonstrate a better trade-off between circuit complexity and performance in single-user multiple-input single-output (MISO) systems. Additionally, optimal architectures that are band-connected and stem-connected BD-RISs, which show improved trade-offs in multi-user multiple-input multiple-output (MIMO) systems, have been introduced in \cite{zhou2025novel, wu2025beyond}.

\begin{figure}[t]
    \centering
    \includegraphics[width= 1\linewidth]{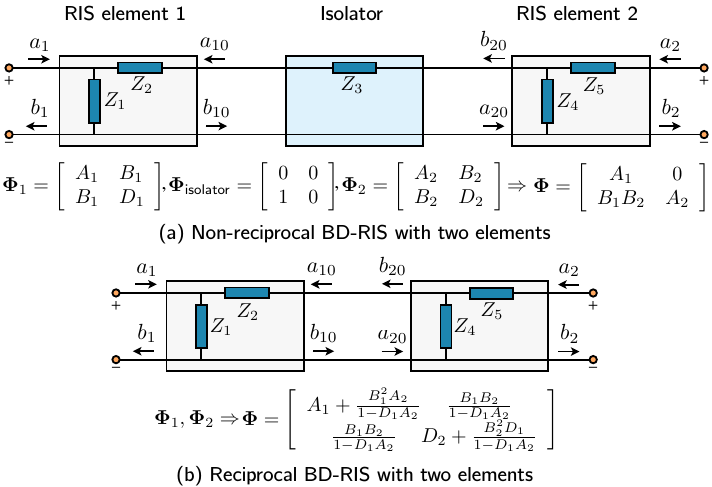}
    \caption{Illustration of 2-port (a) NR-BD-RIS, and (b) R-BD-RIS, where $\boldsymbol{\Phi}_1, \boldsymbol{\Phi}_2$ and $\boldsymbol{\Phi}_\mathsf{isolator}$ are the scattering matrices of the RIS element 1, 2 and isolator, respectively. Based on $\boldsymbol{\Phi}_1, \boldsymbol{\Phi}_2$ and $\boldsymbol{\Phi}_\mathsf{isolator}$, we can obtain the overall scattering matrix $\boldsymbol{\Phi}$ of the $2 \times 2$ BD-RIS.} 
    \label{fig:nr_hardware}
\end{figure}

Circuit reciprocity in the impedance network is a commonly assumed feature in widely-used RIS models, as the impedance between two ports is identical in both directions. This results in a symmetric scattering matrix and channel reciprocity, meaning the signal path from the transmit (Tx) to the receive (Rx) is identical to that from the Rx to the Tx. {While reciprocal RISs, including D-RIS and reciprocal (R)-BD-RIS, have been extensively studied in the literature \cite{shen_modeling_2022, li_reconfigurable_2024, nerini2024beyond, liu2024enhancing}, channel reciprocity imposes limitations on simultaneously serving non-aligned DL and UL devices \cite{li_non-reciprocal_2024}, which is a common scenario in FD LEO satellite systems.} To resolve this constraint, breaking circuit reciprocity is a natural solution, as it disrupts channel reciprocity. Consequently, the reversibility of signal paths is disrupted, enabling support for different signal propagation directions. This capability is particularly beneficial in FD LEO satellite systems, where DL and UL devices are often non-aligned.

Several efforts have been made to realize non-reciprocal (NR)-BD-RIS, such as adding non-reciprocal components \cite{pozar_microwave_2021, xu2024non}, e.g., isolators or circulators, into the impedance network of RIS. In \fig{fig:nr_hardware}, we demonstrate the $2$-port impedance networks and their corresponding scattering matrices of both NR-BD-RIS achieved by isolator and R-BD-RIS \cite{pozar_microwave_2021, xu2024non}. In addition, utilizing radio frequency (RF) micro-electromechanical systems (MEMs) switches \cite{li_reconfigurable_2022, rebeiz2001rf}, and employing metasurface-based techniques \cite{zhang2019breaking, zhang2018space, li2018metasurfaces, liaskos2020internet} can also break the circuit reciprocity. These hardware advancements have enabled NR-BD-RIS to improve channel gain \cite{li_reconfigurable_2022}, achieve spatial selectivity \cite{rusek2024spatially}, and facilitate secure communications \cite{pan_full_duplex_2021, wang2024channel}. A key area of research focuses on supporting non-aligned DL and UL devices in FD systems. For example, \cite{li_non-reciprocal_2024} theoretically demonstrates that NR-BD-RIS enables optimally simultaneous maximization of DL and UL channel gains in single-antenna FD systems, a capability that R-BD-RIS and D-RIS are unable to achieve. Further studies such as \cite{liu2024non} show that  NR-BD-RIS achieves better DL and UL sum-rate performance than other two types of RIS in multi-user multiple-antenna FD systems. Additionally, \cite{liu2025secure} proposes a secure FD wireless circulator, which allows one-way secure communication.
Building on these work, as depicted in \fig{fig:system}, we propose an NR-BD-RIS-enabled FD LEO satellite system with non-aligned Tx and Rx antennas, addressing the needs for low cost, and high spectral efficiency in FD LEO satellite systems. \footnote{{Note that this work assumes an ideal RIS model. In practice, hardware impairments constrain the implementation of NR-BD-RIS. These impairments include discrete tuning steps of the impedance and admittance loads, lossy interconnections and admittance components, frequency-dependent impedance, and mutual coupling \cite{li2024stacked, li2025holographic, li2024performance, li2025tutorial}. Such hardware constraints may introduce deviations from the upper bound performance using ideal model.
}}


\bpara{Contributions and Overview of Results.} The contributions of this paper are listed below:
\begin{enumerate}[leftmargin = *,label =$\bullet$]
    \item To achieve high spectral efficiency by FD transmissions, which are challenging to realize using  conventional RIS, we employ NR-BD-RIS in FD LEO satellite systems. {To address the potential imbalance between DL and UL transmissions, we formulate a weighted DL and UL sum-rate maximization problem.} {Additionally, given the severe path loss in satellite communications, the received uplink power is typically low. Thus, the ratio between self-interference (SI) and uplink signal power is high, making SI mitigation crucial. To address this issue, we consider non-aligned Tx and Rx antennas, as illustrated in \fig{fig:system} (b), to realize propagation-domain SI cancellation. Due to the proximity between the Tx/Rx antennas and the NR-BD-RIS, a near-field channel model is adopted. We incorporate a RIS-assisted satellite far-field channel model. To ensure broad coverage, we design a time-sharing scheduling framework where, in each slot, the NR-BD-RIS serves multiple DL and UL devices in part of the coverage area, and over the full period, it covers the entire area. The antenna configurations, channel models, and time-sharing scheduling framework make our work different from \cite{li_non-reciprocal_2024, liu2024non, liu2025secure}.}
        

    \item {To realize the time-sharing scheduling framework with board area coverage and flexibility in adjusting the DL and UL transmissions, we formulate a weighted DL and UL sum-rate maximization problem over the entire scheduling period.} The optimization variables include the scattering matrix and the Tx and Rx beamforming matrices. To solve this problem, 
    we design an iterative method that updates these variables using a block coordinate descent (BCD) combined with the penalty dual decomposition (PDD) method. Unlike prior works on NR-BD-RIS in \cite{liu2024non, liu2025secure}, the proposed algorithm is designed to the time-sharing scheduling framework.

    \item Numerical results demonstrate that NR-BD-RIS outperforms both R-BD-RIS and D-RIS regarding DL and UL sum-rate performance in single-user (SU) and multi-user (MU) FD LEO satellite systems, even when the Tx and Rx antennas are not perfectly aligned. Furthermore, by analyzing the average DL and UL sum-rate, we show that NR-BD-RIS requires less frequent reconfiguration compared to other two types of RIS, reducing the burden of real-time impedance reconfiguration and enhancing the practicality of the proposed approach. Additionally, the DL and UL weighted sum-rate performance of NR-BD-RIS improves as the number of RIS elements and the group size of BD-RIS increase. We attribute the advantages of NR-BD-RIS to the additional degrees of freedom due to non-reciprocity, enabling simultaneous support for multiple beam directions and non-aligned DL and UL devices. Moreover, the robustness of NR-BD-RIS to SI is analyzed, and the results show that it performs better than R-BD-RIS and D-RIS when the SI power level is below $45$ dB.
\end{enumerate}


\bpara{Organization.} The organization of this paper is given below. We present the system and channel models in Section \ref{sec:system1}. In \ref{sec:problem}, we formulate the optimization problem and develop the proposed solution. Numerical evaluations are provided in \ref{sec:results}. Lastly, we conclude this work in \ref{sec:conclusion}.

\bpara{Notation.} 
We use $\mathbb{R}$ and $\mathbb{C}$ to denote the sets of real and complex numbers. Matrices are written in bold uppercase letters, vectors in bold lowercase letters, and scalar quantities in regular font. The operator $\Re \cdot$ extracts the real part of a complex quantity. For any matrix $\mathbf{X}$, the symbols $\mathbf{X}^*$, $\mathbf{X}^\top$, $\mathbf{X}^H$, and $\mathbf{X}^{-1}$ denote its conjugate, transpose, Hermitian transpose, and inverse. The entry of $\mathbf{X}$ at position $(i,j)$ is written as $[\mathbf{X}]_{i,j}$. The identity matrix and the all-zero matrix are denoted by $\mathbf{I}$ and $\mathbf{0}$. The operators $\operatorname{vec}(\cdot)$, $\operatorname{diag}(\cdot)$, $\operatorname{blkdiag}(\cdot)$, $\Tr(\cdot)$, and $\otimes$ correspond to vectorization, diagonal construction, block-diagonal construction, trace, and the Kronecker product.

\section{System Model}
\label{sec:system1}
\begin{figure*}
    \centering
    \includegraphics[width=0.85\linewidth]{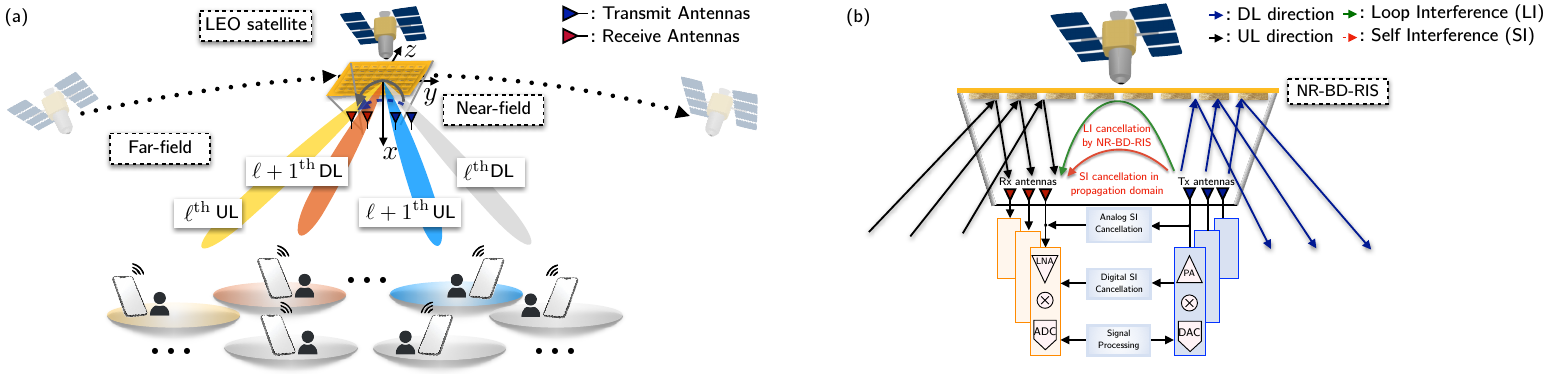}
    \caption{(a) System model of a FD LEO satellite communication system enabled by the NR-BD-RIS. (b) The non-aligned transmit and receive antennas with the NR-BD-RIS
on-board the LEO satellite.}
    \label{fig:system}
\end{figure*}
As shown in \fig{fig:system} (a), we consider a FD LEO satellite system enabled by NR-BD-RIS. The LEO satellite is equipped with $N$ Tx and Rx antennas. {The Tx and Rx antenna are set to be non-aligned as shown in \fig{fig:system}(b)}, which is beneficial to reduce the SI in the propagation domain. The $K$ DL single-antenna devices are indexed by $\mathcal{K}\triangleq\{1, \cdots, K \}$ and $I$ UL single-antenna devices are indexed by $\mathcal{I}\triangleq\{1, \cdots, I \}$. 
The coverage for all the DL/UL devices is achieved through time-sharing beam control. At the $\ell^\mathrm{th}$ time slot, where $\ell \in \mathcal{L} \triangleq \{1, \cdots, L\}$ denotes the time index, the FD LEO satellite serves one group of DL devices with different transmitted data and one group of UL devices with different uploaded data. 

Specifically, the DL and UL devices are both grouped into $L$ groups. The group of DL devices is indexed by $p$, where $p \in \mathcal{L}$. The set of the $p^\mathrm{th}$ DL group is denoted by $\mathcal{G}_{d,p}$. Similarly, the group of UL devices is indexed by $q$, where $q \in \mathcal{L}$. The set of the $q^\mathrm{th}$ UL group is represented by $\mathcal{G}_{u,q}$. Each DL/UL device only belongs to one group, i.e., $\bigcup_{p \in \mathcal{L}} \mathcal{G}_{d,p}=\mathcal{K}$ and $\mathcal{G}_{d,p} \cap \mathcal{G}_{d,m}=\emptyset, \quad \forall p,m \in \mathcal{L}$ and $p \neq m$ for DL devices. $\bigcup_{q \in \mathcal{L}} \mathcal{G}_{u,q}=\mathcal{I}$ and $\mathcal{G}_{u,q} \cap \mathcal{G}_{u,n}=\emptyset, \quad \forall q,n \in \mathcal{L}$ and $q \neq n$ for UL devices. The size of each DL and UL group is defined by $G_{d,p} = | \mathcal{G}_{d,p} |$ and $G_{u,q} = | \mathcal{G}_{u,q} |$, respectively. We also define $\mu_d: \mathcal{K} \rightarrow \mathcal{L}$ and $\mu_u: \mathcal{I} \rightarrow \mathcal{L}$ to map the DL and UL devices to its belonging groups, such that $\mu_d(k) = p, \forall k \in \mathcal{G}_{d,p}$ and $\mu_u(i) = q, \forall i \in \mathcal{G}_{u,q}$. 

The RIS is modeled as a uniform planar array (UPA) with $d_v, d_h \in [\frac{\lambda}{10}, \frac{\lambda}{2}]$ spacing along the vertical and horizontal axes, where $\lambda$ is the carrier wavelength \cite{tang_wireless_2021}. The RIS consists of $M = M_v \times M_h$ elements, where $M_v$ and $M_h$ are the number of elements in the vertical and horizontal directions, respectively. The length of the RIS is indicated by $L_v = M_v d_v$, and the width of the RIS is denoted by $L_h = M_h d_h$. The RIS is placed on the plane of the $y$ and $z$ axes, with the reference point set at the center of the RIS. We assume the Tx and Rx antenna arrays are parallel to the y-axis and are uniform linear arrays (ULAs) consisting of $N$ elements with $d_{a} \in [\frac{\lambda}{10}, \frac{\lambda}{2}]$ spacing. We consider the Tx/Rx antenna to be in the near-field (Fresnel) region of the RIS, and the channel between the Tx antenna and the RIS is modeled as a spherical wavefront \cite{sayeed_deconstructing_2002, tian_near-field_2024}. The Rayleigh distance is given by $r_\mathsf{R} = \frac{2 (L_v^2 + L_h^2)}{\lambda}$, and the near-field region refers to the area inside this distance \cite{liaskos2018new}. The BD-RIS has the scattering matrix $\mathbf{\Phi} \in \mathbb{C}^{M \times M}$. The parameter $M_g$ denotes the group size, and the scattering matrix is divided into $G = M/M_g$ groups. The GC-BD-RIS structure can be written in the form
$\mathbf{\Phi} = \operatorname{blkdiag}(\mathbf{\Phi}_1, \cdots, \mathbf{\Phi}_G)$.



\begin{figure}
    \centering
    \includegraphics[width=0.9\linewidth]{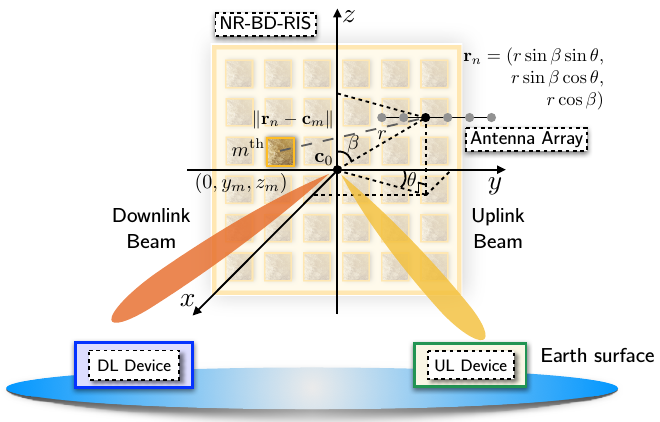}
    \caption{The coordinate system of the BD-RIS channel model. The NR-BD-RIS can support non-aligned downlink and uplink devices simultaneously. }
    \label{fig:nf}
\end{figure}
\vspace{-5pt}

\subsection{Channel Model}
\subsubsection{Near-field Channel Model} 
We consider the channel between the Tx/Rx antenna and the RIS to be in the near-field region due to the short distance. We also assume that the ground devices each have a single antenna. The channel is modeled using a finite-dimensional representation based on the near-field channel model \cite{sayeed_deconstructing_2002, tian_near-field_2024}. The parameters of each path in the channel model are set to be fixed during the channel coherence time. As depicted in \fig{fig:nf}, the RIS is attached to the plane of the $y$ and $z$ axes, with the reference point set at the center of the RIS. The indexing of the RIS elements starts from the bottom-left corner, proceeding in a row by row order. Specifically, the $m^\mathrm{th}$ element is indexed as $m = (m_h - 1) M_v + m_v$, where $m_h \in \mathcal{M}_h \triangleq [1, \cdots, M_h]$ and $m_v \in \mathcal{M}_v \triangleq [1, \cdots, M_v]$ denote the horizontal and vertical indices of the element, respectively. Additionally, $m_h=\left\lfloor\frac{m-1}{M_v}\right\rfloor+1, \quad m_v=\bmod \left(m-1, M_v\right)+1$. The Cartesian coordinates of the $m^\mathrm{th}$ RIS element are given by 
\begin{equation}
    \mathbf{c}_m = [0, (m_h - \frac{M_h + 1}{2})d_h, (m_v - \frac{M_v + 1}{2})d_v]^\top.
\end{equation}

We assume the Tx antenna array is a ULA oriented parallel to the y-axis. The index of the Tx antenna array starts from the left. The reference point of the Tx antenna is located at its center, and the position of the reference point can be described in the polar coordinate system as $\mathbf{p} =[\theta, \beta, r]^\top$. $\theta$ is the azimuth angle, $\beta$ is the elevation angle, and $r$ is the distance between the reference point of Tx antenna and the origin of the coordinate system, \ie $\mathbf{c}_0 = [0, 0, 0]^\top$. The Cartesian coordinates of the $n^\mathrm{th}$ Tx antenna can be expressed as 
\begin{equation}
\begin{aligned}
    \mathbf{r}_n = & [r \sin\beta \sin \theta, r \sin\beta \cos \theta - \frac{(N-1) d_a}{2} + (n-1) d_a, \\
    & r \cos\beta]^\top.
\end{aligned}
\end{equation} 
Thus, under the spherical wavefront model, the channel from the $n^\mathrm{th}$ Tx/Rx antenna element to the $m^\mathrm{th}$ RIS element can be written as
\begin{equation}
[\mathbf{G}]_{m,n} = g_{m,n}  e^{-\jmath \frac{2 \pi}{\lambda} \left\|\mathbf{r}_n-\mathbf{c}_m\right\|},
\end{equation}
where $g_{m,n} = \frac{1}{{4 \pi \left\|\mathbf{r}_n-\mathbf{c}_m\right\|}}$ is the free-space path-loss coefficient between the $n^\mathrm{th}$ Tx/Rx antenna element and the $m^\mathrm{th}$ RIS element \cite{zhang2022beam}, and  $\left\|\mathbf{r}_n-\mathbf{c}_m\right\| $  is the Euclidean distance between the Tx antenna and the $m^\mathrm{th}$ element of the RIS.

\subsubsection{Satellite Communication Channel Model}
The satellite communication channel model is adopted between the BD-RIS and the device. Satellite communications generally have line-of-sight (LoS) propagation. Therefore, the Rician fading model is utilized in satellite communications \cite{tekbiyik_energy-efficient_2022}, given by
\begin{equation}
\mathbf{h} = g_\mathsf{sat} \sqrt{\frac{\kappa}{\kappa+1}} \mathbf{h}_{\mathsf{LoS}} + g_\mathsf{sat} \sqrt{\frac{1}{\kappa+1}} \mathbf{h}_{\mathsf{NLoS}},
\end{equation}
where $\kappa$ is the Rician factor, and $g_\mathsf{sat} = {(PL_\mathsf{BF} PL_\mathsf{rain})}^{-1}$ is the path loss coefficient consisting of free space loss, radiation pattern, and rain attenuation.

\bpara{Path Loss Model.} Taking the channel from the RIS to the ground device as an example, if the Tx antenna is in the near-field region of the RIS, the path loss model used for the near-field beamforming scheme \cite{tang_wireless_2021, tekbiyik_energy-efficient_2022} in an RIS-assisted satellite system is written as:
\begin{equation}
P L_{\mathsf{BF}}=\frac{64 \pi^3}{G_t G_r d_v d_h \lambda^\eta A^\eta \left|\sum_{m=1}^M \frac{\sqrt{F_m^{\mathsf{cmb}}}}{r_{t x} r_{r x}}\right|^\eta},
\end{equation}
where $G_t$ and $G_r$ are the antenna gains of the transmitter and receiver, respectively. $\eta$ denotes the path loss exponent, and $A$ is the reflection coefficient amplitude of each RIS element. The distances from the transmitter to the RIS and from the RIS to the receiver are represented by $r_{tx}$ and $r_{rx}$, respectively. Specifically, $r_{rx}$ is calculated as:
\begin{equation}
r_{rx} = -r_e \sin (\theta_{e})+\sqrt{r_e^2 \sin ^2(\theta_{e})+d_{\mathsf{sat}}^2+2 r_e d_{\mathsf{sat}}},
\end{equation}
where $r_e$ is the radius of the Earth, $d_{\mathsf{sat}}$ is the height of the satellite above the Earth, and $\theta_{e}$ is the elevation angle (\ie the angle with respect to (w.r.t) the positive axis) between the RIS and the ground device. $F_m^{\mathsf {cmb }}$ is the normalized power radiation pattern of the received signal power, which is given by
\begin{equation}
    \label{eq:combine}
F_m^{\mathsf{cmb}}\!=\!F^\mathsf{t x} \! \left(\vartheta_m^\mathsf{t x}, \gamma_m^\mathsf{t x}\right) \! F \!\left(\vartheta_m^\mathsf{t}, \gamma_m^\mathsf{t}\right)\! F\!\left(\vartheta_m^\mathsf{r}, \gamma_m^\mathsf{r}\right)\! F^\mathsf{r x}\! \left(\vartheta_m^\mathsf{rx}, \gamma_n^\mathsf{r x}\right) \!,
\end{equation}
where $F(\vartheta, \gamma)$ is the normalized radiation pattern for the elevation angle $\vartheta$ (\ie the angle w.r.t the positive axis) and the azimuth angle $\gamma$. $F(\vartheta, \gamma)$ is defined as \cite{stutzman2012antenna}
\begin{equation}
F(\vartheta, \gamma)= \begin{cases}\cos ^3 \vartheta & \vartheta \in\left[0, \frac{\pi}{2}\right], \gamma \in[0,2 \pi] \\ 0 & \vartheta \in\left(\frac{\pi}{2}, \pi\right], \gamma \in[0,2 \pi]\end{cases}.
\end{equation}
The angles $\vartheta_m^\mathsf{tx}$, $\gamma_m^\mathsf{tx}$, and $\vartheta_m^\mathsf{rx}$, $\gamma_m^\mathsf{rx}$ denote the elevation and azimuth angles from the transmitting and receiving antennas to the $m^\mathrm{th}$ RIS element, respectively. Similarly, $\vartheta_m^\mathsf{t}$, $\gamma_m^\mathsf{t}$, and $\vartheta_m^\mathsf{r}$, $\gamma_m^\mathsf{r}$ represent the elevation and azimuth angles from the $m^\mathrm{th}$ RIS element to the transmitter and receiver, respectively.

\bpara{Rain attenuation.}
Rain plays a critical role in influencing satellite communication channels, as it leads to both scattering and absorption of electromagnetic waves as they travel through the atmosphere. The rain attenuation is modeled as \cite{itu_p618_13_2017}
\begin{equation}
P L_{\mathsf{rain }}=\xi_R L_E\, (\text{dB}),
\end{equation}
where $\xi_R$ and $L_E$ denote the specific frequency-dependent coefficient \cite{itu_p838_3_2005} and the effective path length.

\subsection{Signal Model}
During a single time slot $\ell$, the FD LEO system simultaneously serves multiple DL and UL devices. For the $k^\mathrm{th}$ DL device, the transmitted signal $s_k$ is sent at time index $\ell$. The effective channel between the Tx antenna and the $k^\mathrm{th}$ DL device is defined as $\widetilde{\mathbf{h}}_{\mathsf{d}, k}^\top \triangleq \mathbf{h}_{\mathsf{d}, k}^\top \mathbf{\Phi}\mathbf{G}_d \in \mathbb{C}^{1 \times N}$, where $\mathbf{h}_{\mathsf{d}, k} \in \mathbb{C}^{M \times 1}$ represents the channel from the BD-RIS to the $k^\mathrm{th}$ DL device, and $\mathbf{G}_d \in \mathbb{C}^{M \times N}$ denotes the channel between the BD-RIS and the Tx antenna. Consequently, the received signal at the $k^\mathrm{th}$ DL device is given by:
\begin{equation}
    \begin{aligned}
    y_{k} = \widetilde{\mathbf{h}}_{\mathsf{d}, k}^\top \mathbf{p}_k s_k + \underbrace{\widetilde{\mathbf{h}}_{\mathsf{d}, k}^\top \sum_{j \in \mathcal{G}_{d, p}, j \neq k} \mathbf{p}_j s_j}_{\text{MU Interference}} +n_k, \quad \forall k \in \mathcal{K},
    \end{aligned}
    \label{eq:dlsignal}
\end{equation}
where $p = \mu_d(k)$ indicates that the $k^\mathrm{th}$ DL device belongs to the $p^\mathrm{th}$ group, and $n_k \sim \mathcal{CN}(0, \sigma^2)$ represents the additive white Gaussian noise (AWGN) at the $k^\mathrm{th}$ DL device.

In the UL direction, the Rx antenna receives the signal, $x_j$, from the $i^{\mathrm{th}}$ UL device, which is reflected by the BD-RIS. Additionally, the Rx antenna receives SI due to coupling between the Tx and Rx \cite{liu2024full}, as well as loop interference caused by the BD-RIS reflecting the transmitted signal. We also define the effective channel between $i^{\mathrm{th}}$ UL user and Rx antenna as $\widetilde{\mathbf{h}}_{\mathsf{u},i} \triangleq  \mathbf{G}_u^\top \mathbf{\Phi} \mathbf{h}_{\mathsf{u}, i} \in \mathbb{C}^{N \times 1}$, where $\mathbf{G}_u \in \mathbb{C}^{M \times N}$ is the channel between BD-RIS and Rx antenna. The received signal are subsequently processed using a combiner $\mathbf{W} \triangleq [\mathbf{w}_1, \dots, \mathbf{w}_{I}] \in \mathbb{C}^{N \times I}$. Hence, the received signal at the Rx antenna from the $i^{\mathrm{th}}$ UL device in $q^\mathrm{th}$ group ($q = \mu_u(i)$) is expressed as:
\begin{equation}
    \begin{aligned}
         z_{i} &
         = \sqrt{P_u} \mathbf{w}_i^H \widetilde{\mathbf{h}}_{\mathsf{u},i} x_i \! +  \! \underbrace{\sqrt{P_u} \sum_{v \in \mathcal{G}_{u,q}, v \neq i }  \mathbf{w}_i^H \widetilde{\mathbf{h}}_{\mathsf{u},v} x_v}_{\text{MU Interference}} \\
        &+ \underbrace{ \mathbf{w}_i^H \mathbf{H}_\mathsf{SI} \mathbf{P} \mathbf{s}}_{\text{Self-interference}} + \underbrace{\mathbf{w}_i^H \mathbf{G}_u^\top \mathbf{\Phi} \mathbf{G}_d \mathbf{P} \mathbf{s}}_{\text{Loop interference}} + \mathbf{w}_i^H \mathbf{n}_u, \quad \forall i \in \mathcal{I},
    \end{aligned}
    \label{eq:rxsignal}
\end{equation}
where $\mathbf{H}_\mathsf{SI} \in  \mathbb{C}^{N \times N}$ denotes the SI channel, and $\mathbf{n}_u \sim \mathcal{CN}(\mathbf{0}, \sigma^2 \mathbf{I})$ is the AWGN at the Rx.

Based on the above signal model, the signal-to-interference-plus-noise ratio (SINR) for the $k^{\mathrm{th}}$ DL user and the SINR at the FD BS for the $i^{\mathrm{th}}$ UL user are given as
\begin{equation}
    \gamma_{\mathsf{d}, k}=\frac{\left|\widetilde{\mathbf{h}}_{\mathsf{d}, k}^\top \mathbf{p}_k\right|^2}{
    \sum_{j \in \mathcal{G}_{d, p}, j \neq k}\left|\widetilde{\mathbf{h}}_{\mathsf{d}, k}^\top \mathbf{p}_j\right|^2+\sigma^2},
    \label{eq:sinr_dl}
\end{equation}
\begin{equation}
    \gamma_{\mathsf{u}, i} =\frac{P_u \left|\mathbf{w}_i^H \widetilde{\mathbf{h}}_{\mathsf{u}, i} \right|^2}{ I_{\mathsf{u}}(\mathbf{P},  \mathbf{W}, \mathbf{\Phi}) + \| \mathbf{w}_i \|^2_F \sigma^2},
\label{eq:sinr_ul}
\end{equation}
where the interference power term regarding the UL transmissions is given by
\begin{equation}
    \begin{aligned}
    I_{\mathsf{u}}&(\mathbf{P},\mathbf{W},  \mathbf{\Phi}) =  P_u \sum_{v \in \mathcal{G}_{u,q}, v \neq i }\left|\mathbf{w}_i^H \widetilde{\mathbf{h}}_{\mathsf{u}, v} \right|^2\\
    &+ \sum_{k \in \mathcal{G}_{d, p}} | \mathbf{w}_i^H \mathbf{H}_\mathsf{SI} \mathbf{p}_k + \mathbf{w}_i^H \mathbf{G}_u^\top \mathbf{\Phi} \mathbf{G}_d \mathbf{p}_k  |^2.
    \end{aligned}
\end{equation}

\begin{remark}
\label{remark1}
We assume perfect instantaneous CSI at the LEO satellite, which can be facilitated by leveraging existing BD-RIS channel estimation methods \cite{li_channel_2024}. Specifically, during the NR-BD-RIS setup phase, the RIS operates in a training mode with predefined scattering matrices. The LEO satellite first transmits pilot signals to the DL user, which estimates the effective DL channel using an LS-based method. The DL user then feeds back the estimated channel coefficients to the satellite via a low-rate control link. Subsequently, the UL user transmits its pilot signals, allowing the LEO satellite to estimate the UL channel.
When multiple $L$ time slots are considered within a scheduling period, the training phase described above is repeated at the beginning of each scheduling period. Due to line-of-sight (LoS) paths dominate in satellite channels, for the remaining $L$ time slots, a geometry-aided prediction approach (\eg Kalman filtering) is employed to estimate the CSI based on the known satellite trajectory and the previously acquired CSI \cite{kim2020massive}. Once the channel estimation process is complete, the NR-BD-RIS transitions to its transmission mode to support data communication efficiently.
\end{remark}

\section{Problem Formulation and Solution}
\label{sec:problem}
\subsection{Weighted DL and UL Sum-rate Maximization over Time}
We aim to maximize the weighted DL and UL sum-rate in the NR-BD-RIS-assisted FD LEO system over the entire scheduling period $L$. In the $\ell^{\mathrm{th}}$ time slot, the system serves $p^\mathrm{th}$ group of DL devices and $q^\mathrm{th}$ group of UL device. Specifically, we optimize the precoding matrix $\mathbf{P}$, combining matrix $\mathbf{W}$, and scattering matrix $\mathbf{\Phi}$. The objective function is defined by:
\begin{equation}
    f_o(\mathbf{\Phi}) \! \triangleq \! \alpha_\mathsf{d} \! \sum_{p \in \mathcal{L}}\sum_{i \in \mathcal{G}_{d,p}} \log_2 (1+\gamma_{\mathsf{d}, k})  +  \alpha_\mathsf{u} \! \sum_{q \in \mathcal{L}}\sum_{j \in \mathcal{G}_{u,q}} \log_2 (1+\gamma_{\mathsf{u}, i}),
\end{equation}
where $\alpha_\mathsf{d}$ and $\alpha_\mathsf{u}$ represent the priorities assigned to DL and UL communications, respectively, with $\alpha_\mathsf{d}+\alpha_\mathsf{u} = 1$. Therefore, we have
\begin{maxi!}|s|[2]                   
    {\mathbf{P}, \mathbf{W}, \mathbf{\Phi}}                               
    {f_o(\mathbf{P}, \mathbf{W}, \mathbf{\Phi}) \label{eq:op1}}   
    {\label{eq:p1}}             
    {\mathcal{P}1:} 
    {} 
    \addConstraint{\|\mathbf{P}\|^2_F}{\leq P_d, \label{eq:op1c1}}
    \addConstraint{\|\mathbf{W}\|^2_F}{= 1, \label{eq:op1c2}}
    \addConstraint{\mathbf{\Phi}}{\in\mathcal{R}_i, \, i \in\{1,2\}, \label{eq:op1c4}}
    \addConstraint{\mathbf{\Phi}}{\in \mathcal{S}_\ell, \, \ell \in\{1,2,3\}. \label{eq:op1c3}}        
\end{maxi!}
Constraint \eqref{eq:op1c4} enforces symmetry for the scattering matrix in the case of R-BD-RIS, \ie $\mathbf{\Phi}\in\mathcal{R}_1 = \{\mathbf{\Phi}  |  \mathbf{\Phi} = \mathbf{\Phi}^\mathsf{T}\}$, and asymmetry for NR-BD-RIS, \ie $\mathbf{\Phi}\in\mathcal{R}_2 = \{\mathbf{\Phi}  | \mathbf{\Phi} \ne \mathbf{\Phi}^\mathsf{T}\}$. Constraint \eqref{eq:op1c3} guarantees the impedance network remains lossless \cite{shen_modeling_2022, pozar_microwave_2021}, with $\mathbf{\Phi}$ satisfying: \textit{i}) $\mathcal{S}_1 = \{ \mathbf{\Phi} = \text{diag}(\phi_1, \ldots,\phi_M) | |\phi_m | = 1, \forall m \in \mathcal{M} \}$ for D-RIS, \textit{ii}) $\mathcal{S}_2 = \{\mathbf{\Phi} = \operatorname{blkdiag}(\mathbf{\Phi}_1, \cdots, \mathbf{\Phi}_G) | \boldsymbol{\Phi}_{g}^H \boldsymbol{\Phi}_{g} = \mathbf{I}, \forall g \in \mathcal{G} \}$ for GC-BD-RIS, and \textit{iii}) $\mathcal{S}_3 = \{\mathbf{\Phi} |  \mathbf{\Phi}^H \mathbf{\Phi} = \mathbf{I} \}$ for fully-connected BD-RIS. Since $\mathcal{P}1$ involves a non-convex objective function with a logarithmic term and fractional structure, along with challenging constraints, a fractional programming approach \cite{shen2018fractional1} is adopted to address it.

\begin{figure*}
\begin{equation}
\begin{aligned}
    &f_\tau(\mathbf{P},\mathbf{W},\mathbf{\Phi},\boldsymbol{\iota}, \boldsymbol{\tau})  =  \alpha_\mathsf{d}  \sum_{p \in \mathcal{L}}\sum_{k \in \mathcal{G}_{d,p}}  \Big( \log_2(1+\iota_{\mathsf{d}, k}) -  \iota_{\mathsf{d}, k} + {2\sqrt{1+\iota_{\mathsf{d}, k}} \Re{\tau_{\mathsf{d}, k}^* \widetilde{\mathbf{h}}_{\mathsf{d}, k}^\top \mathbf{p}_k}}  -  |\tau_{\mathsf{d}, k}|^2 (\sum_{j \in \mathcal{G}_{d, p}}\left|\widetilde{\mathbf{h}}_{\mathsf{d}, k}^\top \mathbf{p}_j\right|^2 + \sigma^2)  \Big)\\ 
    & + \alpha_\mathsf{u} \sum_{q \in \mathcal{L}}\sum_{i \in \mathcal{G}_{u,q}} \Big(  \log_2(1+\iota_{\mathsf{u}, i}) - \iota_{\mathsf{u}, i}  + \! 2 \sqrt{(1+\iota_{\mathsf{u}, i}) P_u}   \Re{ \tau_{\mathsf{u}, i}^* \mathbf{w}_i^H {\widetilde{\mathbf{h}}}_{\mathsf{u}, i}}   -   |\tau_{\mathsf{u}, i}|^2 (I_{\mathsf{u}}(\mathbf{P},  \mathbf{W}, \mathbf{\Phi}) + P_u |\mathbf{w}_i^H {\widetilde{\mathbf{h}}}_{\mathsf{u}, i}|^2  +  \sigma^2)  \Big),
\end{aligned}
\label{eq:ftau}
\end{equation}
\hrulefill
\end{figure*}

\subsection{Problem Transformation}
We reformulate the problem $\mathcal{P}1$ to a more manageable form using fractional programming, which involves two key steps: the Lagrangian dual transformation and the quadratic transformation \cite{shen2018fractional1}. As a result, $\mathcal{P}1$ is transformed into
\begin{maxi!}|s|[2]                   
    {\substack{\mathbf{P}, \mathbf{W},\mathbf{\Phi}, \\ \boldsymbol{\iota}, \boldsymbol{\tau}}}                               
    {f_\tau(\mathbf{P}, \mathbf{W}, \mathbf{\Phi},\boldsymbol{\iota}, \boldsymbol{\tau}) \label{eq:op2}}   
    {\label{eq:p2}}             
    {\mathcal{P}2:} 
    {} 
    \addConstraint{\eqref{eq:op1c1}, \eqref{eq:op1c2}, \eqref{eq:op1c4}, \text{and}\, \eqref{eq:op1c3}}{}
\end{maxi!}
where $f_\tau(\mathbf{\Phi},\boldsymbol{\iota}, \boldsymbol{\tau})$ is defined in \eqref{eq:ftau}. Here, $\boldsymbol{\iota} \triangleq [\boldsymbol{\iota}_{\mathsf{d}}^\top, \boldsymbol{\iota}_{\mathsf{u}}^\top]^\top = [\iota_{\mathsf{d}, 1}, \cdots, \iota_{\mathsf{d}, K}, \iota_{\mathsf{u}, 1}, \cdots, \iota_{\mathsf{u}, I}]^\top \in \mathbb{R}^{K+I}$ and $\boldsymbol{\tau} \triangleq [\boldsymbol{\tau}_{\mathsf{d}}^\top, \boldsymbol{\tau}_{\mathsf{u}}^\top]^\top = [\tau_{\mathsf{d}, 1}, \cdots, \tau_{\mathsf{d}, K}, \tau_{\mathsf{u}, 1}, \cdots, \tau_{\mathsf{u}, I}]^\top \in \mathbb{R}^{K+I}$ are introduced auxiliary variables. To address the constraint for unitarity, the problem is reformulated using the PDD method. The resulting multi-variable optimization is then solved iteratively under the BCD framework, where each variable is updated sequentially until the objective function stabilizes.

The closed-form solutions of the auxiliary vectors $\boldsymbol{\iota}$ and $\boldsymbol{\tau}$ are derived in \cite{shen2018fractional1,liu2024non}. Specifically, the optimal solution of $\boldsymbol{\iota}$ is $\iota_{\mathsf{d}, i}^\mathsf{opt} = \gamma_{\mathsf{d}, i}$ and $\iota_{\mathsf{u}, j}^\mathsf{opt} =\gamma_{\mathsf{u}, j}$. In addition, $\boldsymbol{\tau}$ has the optimal solution given below
\begin{equation}
\tau_{\mathsf{d}, k}^\mathsf{opt} = \frac{ \sqrt{1+\iota_{\mathsf{d}, k}}{\widetilde{\mathbf{h}}}_{\mathsf{d}, k}^\top \mathbf{p}_k}
{\sum_{j \in \mathcal{G}_{d, p}}\left|\widetilde{\mathbf{h}}_{\mathsf{d}, k}^\top \mathbf{p}_j\right|^2
+\sigma^2},
\label{eq:taud}
\end{equation}
\begin{equation}
\tau_{\mathsf{u},i}^\mathsf{opt} = \frac{\sqrt{(1+\iota_{\mathsf{u}, i})} \sqrt{P_u} \mathbf{w}_i^H {\widetilde{\mathbf{h}}}_{\mathsf{u}, i} }
{I_{\mathsf{u}}(\mathbf{P},  \mathbf{W}, \mathbf{\Phi}) + P_u |\mathbf{w}_i^H {\widetilde{\mathbf{h}}}_{\mathsf{u}, i}|^2 + \| \mathbf{w}_i \|^2_F \sigma^2}.
\label{eq:tauu}
\end{equation}

\begin{algorithm}[t]
	\caption{Proposed Algorithm for FD LEO DL and UL Sum-rates over Time Design}
	\label{alg:alg1}
	\KwIn{$\widetilde{\mathbf{H}}_{\mathsf{d}}, \widetilde{\mathbf{H}}_{\mathsf{u}}, \mathbf{H}_\mathsf{SI}, \mathbf{G}_d, \mathbf{G}_u$.}  
	\KwOut{$\mathbf{\Phi}^\mathsf{opt}, \mathbf{P}^\mathsf{opt}, \mathbf{W}^\mathsf{opt}$.} 
	\BlankLine
	Initialize $\mathbf{\Phi}, \mathbf{P}, \mathbf{W}, t=1$.
	
	\While{\textnormal{no convergence of objective function \eqref{eq:op2} \textbf{\&}} $\quad t<t_\mathsf{max}$ }{
        Update $\boldsymbol{\iota}_\mathsf{d}^\mathsf{opt}$ and $\boldsymbol{\iota}_\mathsf{u}^\mathsf{opt}$ by \eqref{eq:sinr_dl} and \eqref{eq:sinr_ul}, respectively. \\
        Update $\boldsymbol{\tau}_\mathsf{d}^\mathsf{opt}$ and $\boldsymbol{\tau}_\mathsf{u}^\mathsf{opt}$ by \eqref{eq:taud} and \eqref{eq:tauu}, respectively. \\
        Update $\mathbf{P}^\mathsf{opt}$ by \eqref{eq:poptimal}. \\
        Update $\mathbf{W}^\mathsf{opt}$ by \eqref{eq:woptimal}. \\
		Update $\mathbf{\Phi}^\mathsf{opt}$ by Algorithm \ref{alg:alg2}. \\
        $t = t + 1$.\\
	}	
	Return $\mathbf{\Phi}^\mathsf{opt}, \mathbf{P}^\mathsf{opt}, \mathbf{W}^\mathsf{opt}$.
\end{algorithm}

\bpara{Transmit Precoder: Block $\mathbf{P}$.} Given that $\mathbf{W}, \mathbf{\Phi},\boldsymbol{\iota}$, and $\boldsymbol{\tau}$ are fixed, we isolate the components associated with $\mathbf{P}$
\begin{equation}
    \begin{aligned}
         &f_\tau(\mathbf{P}) \!  =   \alpha_\mathsf{d} \sum_{p \in \mathcal{L}}\sum_{k \in \mathcal{G}_{d,p}} \Big( \! {2 \sqrt{1+\iota_{\mathsf{d}, k}} \Re{\tau_{\mathsf{d}, k}^* {\widetilde{\mathbf{h}}}_{\mathsf{d}, k}^\top \mathbf{p}_k}} 
        \! \\
        & - \! \mathbf{p}_k^H (\sum_{ w \in \mathcal{G}_{d,p}} |\tau_{\mathsf{d}, k}|^2 {\widetilde{\mathbf{h}}}_{\mathsf{d}, w}^* \widetilde{{\mathbf{h}}}_{\mathsf{d}, w}^\top) \mathbf{p}_k \Big) \! - \!  \alpha_\mathsf{u} \Big(\! \sum_{q \in \mathcal{L}}\sum_{i \in \mathcal{G}_{u,q}}  |\tau_{\mathsf{u}, i}|^2 \\
        & \quad \quad \sum_{k \in \mathcal{G}_{d, p}} |\mathbf{w}_i^H \mathbf{H}_\mathsf{SI} \mathbf{p}_k + \mathbf{w}_i^H \mathbf{G}_u^\top \mathbf{\Phi} \mathbf{G}_d \mathbf{p}_k |^2  \Big).
        \end{aligned}
\end{equation}
Thus, we have the following subproblem
\begin{maxi!}|s|[2]                   
    {\mathbf{P}}                               
    {f_\tau(\mathbf{P}) \label{eq:op3}}   
    {\label{eq:p3}}             
    {\mathcal{P}3:} 
    {} 
    \addConstraint{\|\mathbf{P}\|^2_F}{\leq P_d \label{eq:op3c1}}.
\end{maxi!}
Since both the objective function \eqref{eq:op3} and the constraint \eqref{eq:op3c1} are convex, the Lagrange multiplier method based on the Karush–Kuhn–Tucker (KKT) conditions can be applied to find the optimal solution. By introducing a multiplier $\mu$, the Lagrangian function is formulated as
\begin{maxi!}|s|[2]                   
    {\mathbf{P}}                               
    {f_\tau(\mathbf{P}) - \mu (\|\mathbf{P}\|^2_F - P_d). \label{eq:op4}}   
    {\label{eq:p4}}             
    {\mathcal{P}4:} 
    {} 
\end{maxi!}
Taking the partial derivatives of the Lagrangian function w.r.t $\mathbf{P}$ and $\mu$, and setting them to $\mathbf{0}$, yields the optimal solution for each precoder $\mathbf{p}_k$, as shown in \eqref{eq:poptimal}. The optimal value of $\mu$, denoted as $\mu^\mathsf{opt}$, can be determined using a bisection search.
\begin{figure*}
    \begin{equation}
    \begin{aligned}
        {\mathbf{p}_k^\mathsf{opt}}  = & \Big(
        \alpha_\mathsf{d} \sum_{ w \in \mathcal{G}_{d,p}} \! |\tau_{\mathsf{d}, k}|^2 {\widetilde{\mathbf{h}}}_{\mathsf{d}, w}^* {\widetilde{\mathbf{h}}}_{\mathsf{d}, w}^\top + \alpha_\mathsf{u} \Big( \sum_{i \in \mathcal{G}_{u,q}}   |\tau_{\mathsf{u}, i}|^2 \big( (\mathbf{H}_\mathsf{SI}^H \mathbf{w}_i \mathbf{w}_i^H \mathbf{H}_\mathsf{SI}) + 2 \Re{\mathbf{H}_\mathsf{SI}^H \mathbf{w}_i \mathbf{w}_i^H \mathbf{G}_u^\top \mathbf{\Phi} \mathbf{G}_d}\\
        & + \mathbf{G}_d^H \mathbf{\Phi}^H \mathbf{G}_u^* \mathbf{w}_i \mathbf{w}_i^H \mathbf{G}_u^\top \mathbf{\Phi} \mathbf{G}_d \big) \Big) + \mu^\mathsf{opt} \mathbf{I} \Big)^{-1}  \alpha_\mathsf{d} \sqrt{1+\iota_{\mathsf{d}, k}} \tau_{\mathsf{d}, k} {\mathbf{h}}_{\mathsf{d}, k}^*, \quad k \in \mathcal{K},
    \end{aligned}
    \label{eq:poptimal}
\end{equation}
\hrulefill
\end{figure*}

\bpara{Receive Combiner: Block $\mathbf{W}$.}
With fixed $\mathbf{P}, \mathbf{\Phi},\boldsymbol{\iota}$, and $\boldsymbol{\tau}$, we isolate the terms related to $\mathbf{W}$ in \eqref{eq:ftau}:
\begin{equation}
\begin{aligned}
    f_\tau(\mathbf{W}) &  = \sum_{q \in \mathcal{L}}\sum_{i \in \mathcal{G}_{u,q}}  \Bigg( 2 \sqrt{1+\iota_{\mathsf{u}, i}} \sqrt{P_u} \Re{\tau_{\mathsf{u}, i}^* \mathbf{w}_i^H \widetilde{\mathbf{h}}_{\mathsf{u},i}} \\
    & - | \tau_{\mathsf{u}, i} |^2 \bigg( P_u \sum_{v \in \mathcal{G}_{u,q}} |\mathbf{w}_i^H  \widetilde{\mathbf{h}}_{\mathsf{u},v} |^2 + \sum_{k \in \mathcal{G}_{d,p}} | \mathbf{w}_i^H \mathbf{H}_\mathsf{SI} \mathbf{p}_k\\
    & +\mathbf{w}_i^H \mathbf{G}_{u}^\top \mathbf{\Phi}  \mathbf{G}_d \mathbf{p}_k |^2 + \|\mathbf{w}_i\|^2 \sigma^2 \bigg)
    \Bigg).
\end{aligned}
\end{equation}
We then have the sub-problem:
\begin{maxi!}|s|[2]                   
    {\mathbf{W}}                               
    {f_\tau(\mathbf{W}) \label{eq:op5}}   
    {\label{eq:p5}}             
    {\mathcal{P}5:} 
    {}
    \addConstraint{\|\mathbf{W}\|^2_F=1.}{\label{eq:op5c1}}
\end{maxi!}
Although the constraint \eqref{eq:op5c1} is non-convex, we first treat this sub-problem as an unconstrained complex optimization problem. After convergence, the solution is normalized to satisfy the constraint. The optimal solution for each combiner $\mathbf{w}_i$ is derived by setting $\frac{\partial{f_\tau(\mathbf{W})}}{\partial{\mathbf{w}_i}} = \mathbf{0}$:
\begin{equation}
    \mathbf{w}_i^\mathsf{opt} =  (
     | \tau_{\mathsf{u}, i}|^2 \boldsymbol{\zeta} )^{-1}
    (\sqrt{1+\iota_{\mathsf{u}, i}} \sqrt{P_u} \tau_{\mathsf{u}, i}^* {\widetilde{\mathbf{h}}}_{\mathsf{u},i} ), \, \forall i\in \mathcal{I},
\label{eq:woptimal}
\end{equation}
where $\boldsymbol{\zeta}$ is defined as:
\begin{equation}
\begin{aligned}
    \boldsymbol{\zeta} & = P_u \sum_{v \in \mathcal{G}_{u,q}}  {\widetilde{\mathbf{h}}}_{\mathsf{u},v} \widetilde{{\mathbf{h}}}_{\mathsf{u},v}^H  + \mathbf{H}_\mathsf{SI}  \sum_{ k \in \mathcal{G}_{d,p}} \mathbf{p}_k \mathbf{p}_k^H \mathbf{H}_\mathsf{SI}^H \\
    & + \mathbf{G}_u^\top \mathbf{\Phi} \mathbf{G}_d \sum_{{ k \in \mathcal{G}_{d,p}}} \mathbf{p}_k \mathbf{p}_k^H \mathbf{G}_d^H \mathbf{\Phi} ^H \mathbf{G}_u^* \\
    & + 2 \Re{\mathbf{G}_u^\top \mathbf{\Phi} \mathbf{G}_d \sum_{k \in \mathcal{G}_{d,p}} \mathbf{p}_k \mathbf{p}_k^H \mathbf{H}_\mathsf{SI}^H}
    + \sigma^2.
\end{aligned}
\end{equation}
Finally, $\mathbf{W}$ is normalized (\ie $\mathbf{W}/\| \mathbf{W}\|_F$) to satisfy the receiver-side power constraint.

\begin{algorithm}[t]
	\caption{Algorithm for Updating BD-RIS Scattering Matrix $\mathbf{\Phi}$}
	\label{alg:alg2}
	\KwIn{$\mathbf{P}, \mathbf{W}, \boldsymbol{\iota}, \boldsymbol{\tau},\widetilde{\mathbf{H}}_{\mathsf{d}}, \widetilde{\mathbf{H}}_{\mathsf{u}}, \mathbf{H}_\mathsf{SI}, \mathbf{G}_d, \mathbf{G}_u$.} 
	Initialize $\{\mathbf{\Phi}_g\}, \{\mathbf{\Psi}_g\}, \{\mathbf{\Lambda}_g\}, \rho, t_\mathsf{inner}=t_\mathsf{outer}=1$.
 
    \For{$g \gets 1$ \KwTo $G$}{
        \While{\textnormal{$\| \mathbf{\Phi}_g - \mathbf{\Psi}_g \|_\infty > \varepsilon$ \textbf{\&}} $ t_\mathsf{outer}<t_\mathsf{outer\, max}$ }{
        \While{\textnormal{no convergence of objective function \eqref{eq:op7} \textbf{\&}} $ t_\mathsf{inner}<t_\mathsf{inner\, max}$ }{
        Update $ \mathbf{\Phi}_g $ by \eqref{eq:phi}.\\
        Update $ \mathbf{\Psi}_g $ by \eqref{eq:phi_copy}.\\
        $t_\mathsf{inner} = t_\mathsf{inner} + 1$.
	}
    \uIf{$\| \mathbf{\Phi}_g - \mathbf{\Psi}_g \|_\infty < \epsilon$}{
        $\mathbf{\Lambda}_g = \mathbf{\Lambda}_g + \rho^{-1} (\mathbf{\Phi}_g - \mathbf{\Psi}_g)$.
    }
    \Else{
        $\rho = c\rho$.
    }
    $t_\mathsf{outer} = t_\mathsf{outer} + 1$.
	}	
    }
	\KwRet{ $\mathbf{\Phi}^\mathsf{opt} = \operatorname{blkdiag}(\mathbf{\Phi}^\mathsf{opt}_1, \cdots, \mathbf{\Phi}^\mathsf{opt}_G)$.}
\end{algorithm}

\bpara{BD-RIS Scattering Matrix: $\mathbf{\Phi}$.} 
To update the scattering matrix $\mathbf{\Phi}$, there are two challenges: 1) the coupling between the constraints \eqref{eq:op1c4} and \eqref{eq:op1c3}; 2) the constraint for symmetry, \ie \eqref{eq:op1c4}. To address the first challenge, the PDD method is adopted to decouple the constraints. Subsequently, the second challenge is addressed by linear reformulation of the scattering matrix. The details are given as follows.

We first isolate the terms w.r.t $\boldsymbol{\Phi}$ with fixed $\mathbf{P}, \mathbf{W},\boldsymbol{\iota}$, and $\boldsymbol{\tau}$ given in \eqref{eq:ftau_phi}.
\begin{figure*}
\begin{equation}
    \begin{aligned}
        &f_\tau(\mathbf{\Phi}) = \alpha_\mathsf{d} \sum_{p \in \mathcal{L}}\sum_{k \in \mathcal{G}_{d,p}} 2 \sqrt{1+\iota_{\mathsf{d}, k}} \Re{\tau_{\mathsf{d}, k}^* \mathbf{h}_{\mathsf{d}, k}^\top \mathbf{\Phi} \mathbf{G}_d \mathbf{p}_k}    
        + \alpha_\mathsf{u} \sqrt{P_u} \sum_{q \in \mathcal{L}}\sum_{i \in \mathcal{G}_{u,q}}  2 \sqrt{1+\iota_{\mathsf{u}, i}} \Re{\tau_{\mathsf{u}, i}^* \mathbf{w}_i^H \mathbf{G}_u^\top  \mathbf{\Phi} \mathbf{h}_{ \mathsf{u}, i}}  \\
        & - \alpha_\mathsf{d} \! \sum_{p \in \mathcal{L}}\sum_{k \in \mathcal{G}_{d,p}} \! \!|\tau_{\mathsf{d}, k}|^2 \sum_{j \in \mathcal{G}_{d,p}} |{\widetilde{\mathbf{h}}}_{\mathsf{d}, k}^\top \mathbf{p}_j|^2
        - \alpha_\mathsf{u} \!\! \sum_{q \in \mathcal{L}}\sum_{i \in \mathcal{G}_{u,q}} \! |\tau_{\mathsf{u}, i}|^2 \!  \bigg( \! \!
        P_u \sum_{v \in \mathcal{G}_{u,q}} |\mathbf{w}_i^H  \widetilde{\mathbf{h}}_{\mathsf{u},v} |^2 + \! \! \sum_{k \in \mathcal{G}_{d,p}} | \mathbf{w}_i^H \mathbf{H}_\mathsf{SI} \mathbf{p}_k +\mathbf{w}_i^H \mathbf{G}_{u}^\top \mathbf{\Phi} \mathbf{G}_d \mathbf{p}_k |^2 \!
        \bigg)\\
        & = 2 \alpha_\mathsf{d} \sum_{p \in \mathcal{L}} \Re{ \Tr({\mathbf{C}_1}  \mathbf{\Phi})} 
        + 2 \alpha_\mathsf{u} \sqrt{P_u} \sum_{q \in \mathcal{L}} \Re{\Tr({\mathbf{C}_2}\mathbf{\Phi} )} 
        - \alpha_\mathsf{d} \sum_{p \in \mathcal{L}} \Tr({\mathbf{B}_1} \mathbf{\Phi} {\mathbf{A}_1} \mathbf{\Phi}^H ) 
        - \alpha_\mathsf{u} P_u \sum_{q \in \mathcal{L}} \Tr({\mathbf{A}_2} \mathbf{\Phi} {\mathbf{B}_2}  \mathbf{\Phi}^H) \\
        & \quad - \alpha_\mathsf{u} \sum_{q \in \mathcal{L}} \Tr(  {\mathbf{A}_2} \mathbf{\Phi} {\mathbf{A}_1} \mathbf{\Phi}^H),
    \end{aligned}
    \label{eq:ftau_phi}
\end{equation}
\hrulefill
\end{figure*}
We then employ a two-loop PDD method, introducing a replica $\mathbf{\Psi}$ of $\mathbf{\Phi}$ to decouple constraints \eqref{eq:op1c4} and \eqref{eq:op1c3}, while adding the equality constraint $\mathbf{\Phi} = \mathbf{\Psi}$. An augmented Lagrangian problem $\mathcal{P}3$ is formulated by incorporating the equality constraint into the objective function using a dual variable $\mathbf{\Lambda}$ and a penalty coefficient $\rho^{-1}$. The inner loop focuses on solving the augmented Lagrangian problem $\mathcal{P}3$, while the outer loop updates the dual variable $\mathbf{\Lambda}_g$ and the penalty coefficient $\rho^{-1}$ as needed to ensure the equality constraint \eqref{eq:op70c1} is satisfied.

Specifically, the sub-optimization problem after introducing the copy $\mathbf{\Psi}$ and equality constraint $\mathbf{\Phi} = \mathbf{\Psi}$ is given by
\begin{maxi!}|s|[2]                   
    {\mathbf{\Phi}, \mathbf{\Psi}}                               
    {f_\tau(\mathbf{\Phi}) \label{eq:op70}}   
    {\label{eq:p70}}             
    {\mathcal{P}3:} 
    {}
    \addConstraint{\mathbf{\Psi}_g^H \mathbf{\Psi}_g = \mathbf{I},}{\quad  \forall g \in \mathcal{G} \label{eq:op70c1},}
    \addConstraint{\mathbf{\Phi}_g = \mathbf{\Psi}_g,}{\quad  \forall g \in \mathcal{G} \label{eq:op70c2},}
    \addConstraint{\mathbf{\Phi}_g \in\mathcal{R}_i, \quad i \in\{1,2\}, \, \forall g \in \mathcal{G}.}{\label{eq:op70c3}}
\end{maxi!}
Then, the transformed augmented Lagrangian problem is:
\begin{mini!}|s|[2]                   
    {\mathbf{\Phi}, \mathbf{\Psi}}                               
    {- \!f_\tau(\mathbf{\Phi}) \!+  \!\frac{1}{2 \rho}  \!\sum_{g \in \mathcal{G}}\! \|\mathbf{\Phi}_g \! - \! \mathbf{\Psi}_g \|^2   \label{eq:op7}}   
    {\label{eq:p7}}             
    {\mathcal{P}4:} 
    \breakObjective{+ \sum_{g \in \mathcal{G}} \Re{\Tr(\mathbf{\Lambda}_g^H(\mathbf{\Phi}_g - \mathbf{\Psi}_g))}}{\nonumber}
    {}
    \addConstraint{\eqref{eq:op70c1}, \eqref{eq:op70c3}.}{\nonumber}
\end{mini!}

\begin{figure}[t]
    \centering
    \includegraphics[width=0.35\textwidth]{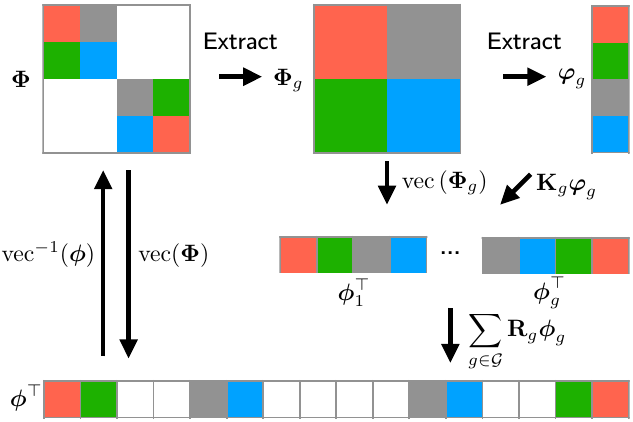}
    \caption{An example of linear reformulation based on non-reciprocal case.}
    \label{fig:linear}
\end{figure}
Next, we adopt linear reformulation of the scattering matrix to address the constraint for symmetry \eqref{eq:op70c3}. The needed elements inside $\mathbf{\Phi}_g$ are extracted in a column by column order and vectorized as $\boldsymbol{\varphi}_g$. Note that the need elements of NR-BD-RIS are all the elements in $\mathbf{\Phi}_g$, while those of R-BD-RIS are the upper/lower triangular elements and the diagonal elements. 

To recover the original matrix from the extracted vector $\boldsymbol{\varphi}_g$, we introduce permutation matrix $\mathbf{K}_{g}$ and mapping matrix $\mathbf{R}_g$ in \cite{liu2024non} and utilize them for the permutation process in \eqref{eq:permutation} and mapping process in \eqref{eq:mapping}, respectively. An illustration of these processes with the non-reciprocal case is given in Fig. \ref{fig:linear}, where the vectorization is $\boldsymbol{\psi}_g \triangleq \operatorname{vec}(\boldsymbol{\Psi}_g)$, $\lambda_g \triangleq \operatorname{vec}(\boldsymbol{\Lambda}_g)$.
\begin{equation}
    \boldsymbol{\phi}_g \triangleq \operatorname{vec}\left(\boldsymbol{\Phi}_g\right)=\mathbf{K}_{g} \boldsymbol{\varphi}_g, \quad \forall g \in \mathcal{G}.
\label{eq:permutation}
\end{equation}
\begin{equation}
\boldsymbol{\phi} \triangleq \operatorname{vec}(\boldsymbol{\Phi})=\sum_{g \in \mathcal{G}} \mathbf{R}_g \boldsymbol{\phi}_g.
\label{eq:mapping}
\end{equation}

Following the preprocessing step, the two-loop PDD method can be applied to solve the problem $\mathcal{P}4$.
\subsubsection{Update of $\mathbf{\Phi}_g$ in the Inner Loop}
The sub-problem w.r.t $\mathbf{\Phi}_g$ in the form of $\boldsymbol{\varphi}_g$ is written as 
\begin{mini!}|s|[2]                   
    {\boldsymbol{\varphi}_g}                               
    {L( \boldsymbol{\varphi}_g ) = \boldsymbol{\varphi}_g^H \boldsymbol{\Delta} \boldsymbol{\varphi}_g - 2 \Re{\boldsymbol{\varphi}_g^H \boldsymbol{\delta}}. \label{eq:op8}}   
    {\label{eq:p8}}             
    {\mathcal{P}5:} 
    {}
\end{mini!}
Since $\mathcal{P}5$ is an unconstrained quadratic optimization problem, the optimal solution is obtained by setting $\frac{\partial{L( \boldsymbol{\varphi}_g )}}{\partial{\boldsymbol{\varphi}_g}} = \mathbf{0}$. Therefore, the optimal solution of $\boldsymbol{\varphi}_g^\mathsf{opt}$ is:
\begin{equation}
    \boldsymbol{\varphi}_g^\mathsf{opt} = \boldsymbol{\Delta}^{-1} \boldsymbol{\delta},
    \label{eq:phi}
\end{equation}
where $\boldsymbol{\Delta}$ and $\boldsymbol{\delta}$ are defined as 
\begin{equation}
\begin{aligned}
     \boldsymbol{\Delta} &  = \alpha_\mathsf{u} \mathbf{K}_{g}^H \mathbf{R}_g^H \sum_{q \in \mathcal{L}} \big(
    P_u \mathbf{B}_2^\top \otimes \mathbf{A}_2 + \mathbf{A}_1^\top \otimes \mathbf{A}_2 \big) \mathbf{R}_g \mathbf{K}_{g}\\
    & + \alpha_\mathsf{d} \mathbf{K}_{g}^H \mathbf{R}_g^H \sum_{p \in \mathcal{L}} \big(\mathbf{A}_1^\top \otimes \mathbf{B}_1 \big) \mathbf{R}_g \mathbf{K}_{g}  + \frac{1}{2 \rho} \mathbf{K}_{g}^H \mathbf{K}_{g},
    \label{eq:bigDelta}
\end{aligned}
\end{equation}
\begin{equation}
\begin{aligned}
     \boldsymbol{\delta}&  =  \alpha_\mathsf{d} \mathbf{K}_{g}^H \mathbf{R}_g^H \sum_{p \in \mathcal{L}} 
    \operatorname{vec}^*(\mathbf{C}_1^\top)  + \! \alpha_\mathsf{u}  \mathbf{K}_{g}^H \mathbf{R}_g^H \sum_{q \in \mathcal{L}} 
    \sqrt{P_u} \operatorname{vec}^*(\mathbf{C}_2^\top) \\  
    & \quad + \frac{1}{2\rho} \mathbf{K}_{g}^H \mathbf{K}_{g} \boldsymbol{\psi}_g - \frac{1}{2} \mathbf{K}_{g}^H \boldsymbol{\lambda}_g.
    \label{eq:smalldelta}
\end{aligned}
\end{equation}
The introduced expressions are listed in the Table \ref{tab:1}. 
\begin{table}[t]
\caption{Newly Introduced Notations}
\resizebox{0.5\textwidth}{!}{
\renewcommand{\arraystretch}{1.8}
\begin{tabular}{|l|l|}
\hline
$\mathbf{A}_1 = \sum_{j \in \mathcal{G}_{d,p}} \mathbf{G}_d \mathbf{p}_j \mathbf{p}_j^H \mathbf{G}_d^H$ &$\mathbf{A}_2 =  \sum_{i \in \mathcal{G}_{u,q}}  |\tau_{\mathsf{u}, i}|^2 \mathbf{G}_u^* \mathbf{w}_i \mathbf{w}_i^H \mathbf{G}_u^\top$   \\ \hline
$ \mathbf{B}_1 = \sum_{k \in \mathcal{G}_{d,p}} |\tau_{\mathsf{d}, k}|^2 \mathbf{h}_{\mathsf{d}, k}^* \mathbf{h}_{\mathsf{d}, k}^\top$  & $\mathbf{B}_2 =  \sum_{v \in \mathcal{G}_{u,q}} \mathbf{h}_{\mathsf{u},v} \mathbf{h}_{\mathsf{u},v}^H$ 
\\ \hline
$\mathbf{C}_1 = \sum_{k \in \mathcal{G}_{d,p}} \sqrt{1+\iota_{\mathsf{d}, k}} \tau_{\mathsf{d}, k}^* \mathbf{G}_d \mathbf{p}_k \mathbf{h}_{\mathsf{d}, k}^\top$ 
& 
$\mathbf{C}_2 = \sum_{i \in \mathcal{G}_{u,q}} \sqrt{1+\iota_{\mathsf{u}, i}} \tau_{\mathsf{u}, i}^* \mathbf{h}_{\mathsf{u},i} \mathbf{w}_i^H \mathbf{G}_u^\top$ 
\\ 
\hline
\end{tabular}
\label{tab:1}
}
\end{table}

\subsubsection{Update of $\mathbf{\Psi}_g$ in the Inner Loop}
The optimization problem involving $\mathbf{\Psi}_g$ is an orthogonal procrustes problem and given by
\begin{mini!}|s|[2]                   
    {\boldsymbol{\Psi}_g}                               
    {\| \boldsymbol{\Psi}_g - (\rho \boldsymbol{\Lambda}_g + \boldsymbol{\Phi_g})  \|^2_F, \label{eq:op9}}   
    {\label{eq:p9}}             
    {\mathcal{P}5:} 
    {}
    \addConstraint{\mathbf{\Psi}_g^H \mathbf{\Psi}_g = \mathbf{I}.}{\label{eq:op9c1}}
\end{mini!}
This problem has a closed-form solution given by \cite{manton2002optimization}
\begin{equation}
\boldsymbol{\Psi}_g^{\mathsf{opt}}=\mathbf{U}_g  \mathbf{V}_g^H,
\label{eq:phi_copy}
\end{equation}
where the unitary matrices $\mathbf{U}_g$ and $\mathbf{V}_g$ are derived from the singular value decomposition (SVD) of $\rho \boldsymbol{\Lambda}_g + \boldsymbol{\Phi}_g$.

\subsubsection{Update of $\mathbf{\Lambda}_g$ and $\rho$ in the Outer Loop}
Once the inner loop converges, the dual variable $\mathbf{\Lambda}_g$ and the penalty coefficient $\rho$ are updated in the outer loop. The convergence criterion for the outer loop is defined as $\| \boldsymbol{\Phi}_g - \boldsymbol{\Psi}_g \|_\infty < \epsilon$, where $\epsilon$ is a small positive threshold. This ensures that the equality constraint \eqref{eq:op70c2} is satisfied. If the criterion is met, the dual variable is updated as
\begin{equation}
    \mathbf{\Lambda}_g = \mathbf{\Lambda}_g + \rho^{-1} (\mathbf{\Phi}_g - \mathbf{\Psi}_g).
\end{equation}
If $\mathbf{\Phi}_g$ and $\mathbf{\Psi}_g$ are not sufficiently close, the penalty coefficient is updated as $\rho = c\rho, c \in (0,1)$. Once the optimal solution for each group $\mathbf{\Phi}_g^\mathsf{opt}$ is obtained, the overall scattering matrix $\mathbf{\Phi}^\mathsf{opt}$ can be reconstructed. The procedure for updating the scattering matrix $\mathbf{\Phi}$ is outlined in Algorithm \ref{alg:alg2}.

\subsection{Convergence Analysis}
\begin{figure}
    \centering
    \includegraphics[width=0.85\linewidth]{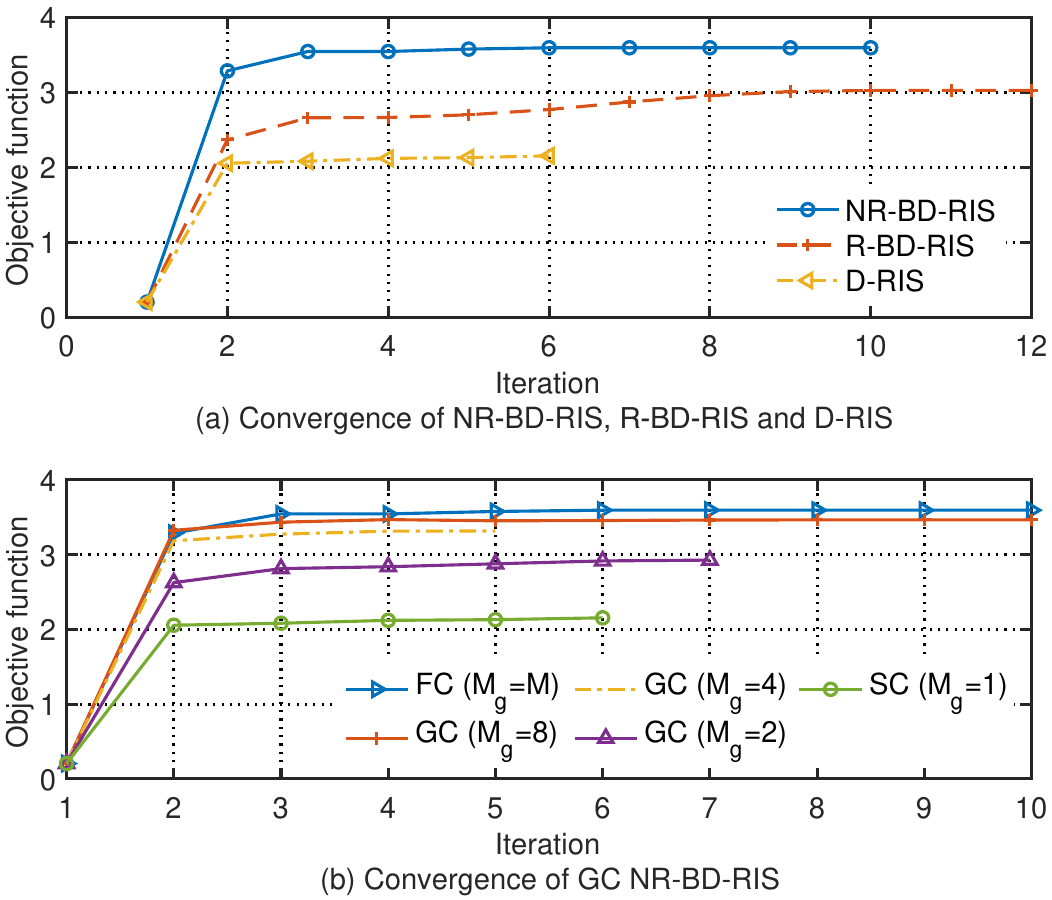}
    \caption{Convergence of the proposed algorithm. The Tx and Rx antenna are fixed at $88^\circ$ and $92^\circ$, respectively. The RIS arrays with $M=16$ RIS elements are set to a UPA ($4 \times 4$). The number of Tx and Rx antennas are both $N=2$, and $G_{d,p} = 2$ DL devices  and $G_{u,q} = 2$ UL devices are served at each time slot. The RIS is configured as a UPA ($4 \times 4$) with $M=16$ elements, and the update interval is set to $L=2$. (a) Convergence for FC NR-BD-RIS, R-BD-RIS, and D-RIS. (b) Convergence for GC NR-BD-RIS with varying group sizes.}
    \label{fig:convergence}
\end{figure}
The convergence of the proposed algorithm is evaluated. The objective function \eqref{eq:op2} is guaranteed to converge to at least a local optimum. Let $v^t$ represent the objective function value at the $t^\mathrm{th}$ iteration. In Algorithm \ref{alg:alg1}, the blocks are updated through an iterative procedure. For each group $g \in \mathcal{G}$, the PDD step used to update $\mathbf{\Phi}_g$ converges within its inner loop, since the related sub-problems are convex and have unique solutions. The outer loop of the PDD method converges to a KKT point as the augmented Lagrangian enforces constraint satisfaction while updating the dual variables \cite{shi2020penalty}. As a result, $\mathbf{\Phi}$ converges to a local optimum. Similarly, the optimization of $\mathbf{\iota}$, $\mathbf{\tau}$, $\mathbf{P}$, and $\mathbf{W}$ is convex, ensuring that the objective function \eqref{eq:op2} is monotonically non-decreasing, \ie $v^t \geq v^{t-1}$. Therefore, the overall algorithm converges to a local optimum.

Simulation results in \fig{fig:convergence} (a) confirm the convergence of Algorithm \ref{alg:alg1} for fully-connected NR-BD-RIS, R-BD-RIS, and D-RIS, with all cases converging within $12$ iterations. The NR-BD-RIS achieves the highest objective value, highlighting its superior performance. \fig{fig:convergence} (b) shows the convergence for GC-NR-BD-RIS, where smaller group sizes $M_g$ lead to lower objective values due to reduced flexibility in the scattering matrix design.


\subsection{Complexity Analysis}
In the BCD framework, each block is optimized iteratively. The update of $\boldsymbol{\iota}$ and $\boldsymbol{\tau}$ has a per-iteration cost of $\mathcal{O}(K^2 M^2)$, since it mainly consists of element-wise computations and matrix–vector products. The design of precoder $\mathbf{P}$ involves matrix inversion together with a bisection search, giving a complexity of $\mathcal{O}(K(M^2 + I_\mathsf{b} N^3))$, where $I_\mathsf{b}$ denotes the number of bisection steps. The combiner $\mathbf{W}$ update also relies on matrix inversion and incurs a cost of $\mathcal{O}(K N^3)$.
Among all blocks, the PDD update of $\mathbf{\Phi}$ requires the largest computational effort. Specifically, the Kronecker product in \eqref{eq:phi} contributes $\mathcal{O}(M_g^2 M^4)$, while the SVD operation in \eqref{eq:phi_copy} adds $\mathcal{O}(M_g^3)$. Consequently, the complexity of Algorithm \ref{alg:alg2} is $\mathcal{O}(G I_\mathsf{outer} I_\mathsf{inner} M_g^2 M^4)$, where $I_\mathsf{outer}$ and $I_\mathsf{inner}$ denote the outer and inner iteration counts in the PDD method, respectively.
Overall, the total complexity of the proposed algorithm is $\mathcal{O}(I_\mathsf{BCD} (K^2 M^2 + K(M^2 + I_\mathsf{b} N^3) + K N^3 + G I_\mathsf{outer} I_\mathsf{inner} M_g^2 M^4))$, where $I_\mathsf{BCD}$ represents the number of BCD iterations needed to reach convergence.

\begin{figure}
    \centering
    \includegraphics[width=0.8\linewidth]{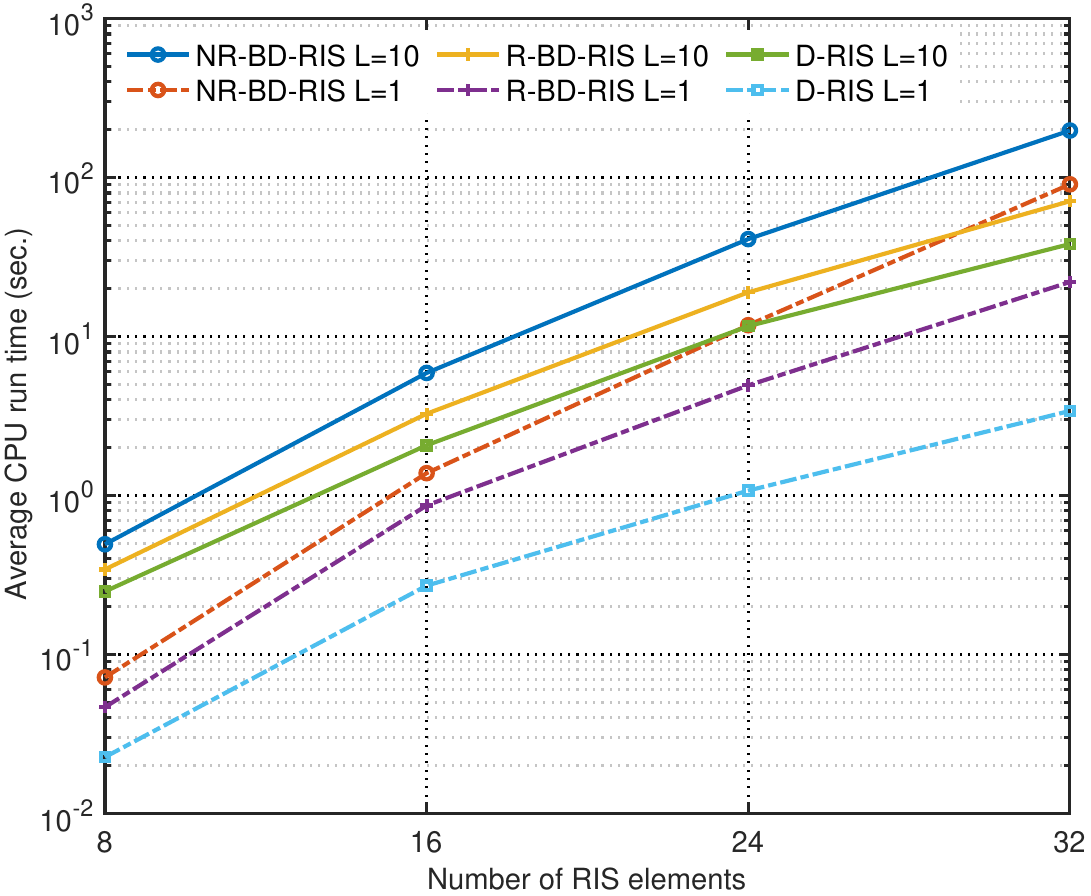}
    \caption{{Average CPU running time for diagonal RIS, reciprocal BD-RIS and non-reciprocal BD-RIS versus the number of RIS elements $M$ with update interval $L=1$ and $10$.}}
    \label{fig:runtime}
\end{figure}

We also evaluate the average CPU running time of the proposed algorithm for different RIS architectures with different update intervals, as shown in \fig{fig:runtime}. {The update interval represents the number of consecutive time slots over which the same RIS configuration is reused. Specifically, an update interval of $1$ implies that the RIS is reconfigured at every time slot, whereas an update interval of $L$ indicates that the RIS is optimized once and applied to successive $L$ time slots.} It has shown that the running time follows polynomial growth with respect to the number of RIS elements $M$, which is consistent with the derived computational complexity. In addition, it can be observed that the running time increases with larger update interval $L$, {since more matrix operations are needed for the optimization of $\mathbf{\Phi}$ for the total $L$ time slots, especially the steps in \eqref{eq:phi}, \eqref{eq:bigDelta}, and \eqref{eq:smalldelta}. }
Notably, the running time for NR-BD-RIS is higher than that of R-BD-RIS and D-RIS. The channel coherence time is assumed to be larger than the running time of the proposed algorithm to ensure real-time implementation.

{To reduce computational complexity, adapt to real-time CSI variations, and improve scalability for larger RIS sizes and more user deployments, future work can build on the projected gradient descent (PGD) method \cite{zhou2025joint}. Specifically, the PGD update applies the projection and thus transforms the double-loop PDD method into a single-loop PGD method, achieving a per-iteration complexity of $\mathcal{O}(M^3)$. Additionally, the partially proximal alternating direction method of multipliers (ADMM) \cite{wu2024optimization} can also enhance scalability, with a per-iteration complexity of $\mathcal{O}(K^3 M^3)$.}

\section{Numerical Evaluation}
\label{sec:results}
This section assesses the performance of the proposed NR-BD-RIS-assisted FD LEO satellite system and compares its DL and UL sum-rates with those of the R-BD-RIS and D-RIS, showing the benefit of the NR-BD-RIS. Specifically, in the SU case, we analyze the impact of the following parameters on the DL and UL sum-rate performance: 1) angles of ground UL devices, 2) angular separation between the Tx and Rx antennas, 3) RIS update interval, 4) sum-rate region by varying DL and UL weights, 5) number of RIS elements, 6) group size of the GC NR-BD-RIS, 7) SI power level, and 8) number of DL and UL devices. Additionally, in the MU case, we assess the performance by examining the impact of RIS update interval and the sum-rate region for different numbers of DL and UL devices.


In the simulation, we consider both SU case (\ie $G_{d,p}=1, G_{u,q}=1$) and MU cases (\ie $G_{d,p} > 1, G_{u,q} > 1$) within each time slot. Since the LoS links  are common in satellite communications, we model the far-field channel between the RIS and ground DL/UL devices using a Rician factor $\kappa = 10$ dB. The DL and UL priority parameters are set equally, \ie $\alpha_d = \alpha_u = 0.5$. To suppress the power level of SI, the Tx and Rx antennas are positioned at $90 \pm \Delta_\mathsf{deg}$, where $\Delta_\mathsf{deg}$ is a small angular deviation from $90^\circ$. Specifically, the Tx and Rx antennas are fixed at $88^\circ$ and $92^\circ$, respectively, ensuring that the incident wave from the Tx antenna to the RIS is nearly orthogonal to the RIS plane (yz plane). This configuration maximizes the normalized power radiation pattern in \eqref{eq:combine}. {Since the Tx and Rx antennas are close to the RIS, the loop interference is treated as reflected SI. We assume that the total SI is pre-suppressed using SI cancellation techniques and is set to $20$ dB larger than the noise floor. Note that the loop interference (reflected SI) can be effectively mitigated using analog and digital cancellation methods with more frequent reconstruction of SI to track the changes of scattering matrix \cite{kolodziej2019band, liu2017full, liu2024full}} Additionally, we consider fully-connected NR-BD-RIS and R-BD-RIS. The simulation parameters, particularly those relevant to the LEO satellite, are summarized in Table \ref{tab:2}.

\begin{table}[t]
  \centering
\caption{Simulation Parameters}
  \begin{tabular}{l|c}
    \toprule[1pt]
    Parameters      &  Value  \\ \hline \hline
    
    Transmit budget $P_t$ &  50 dBm \\ \hline
    Noise power spectral density&  -174 dBm/Hz \\ \hline
    Carrier frequency $f_c$ &  10.5 GHz (K band) \\ \hline
    Satellite height $d_\mathsf{sat}$ &  $600$ km \\ \hline
    Bandwidth $B$ & $1$ MHz  \\ \hline
    Pathloss coefficient $\eta$ & 2  \\ \hline
    Distance between the Tx antenna and the RIS $r_{tx}$ & $5$ m \\ \hline
    Antenna gain for the Tx and Rx $G_t, G_r$ &  $30$, $33.4$ dB \\ \hline
    Rainfall rate & 31.119  \\ \hline
    \bottomrule[1pt]
  \end{tabular}
  \label{tab:2}
\end{table}

\begin{figure}
    \centering
    \includegraphics[width=1\linewidth]{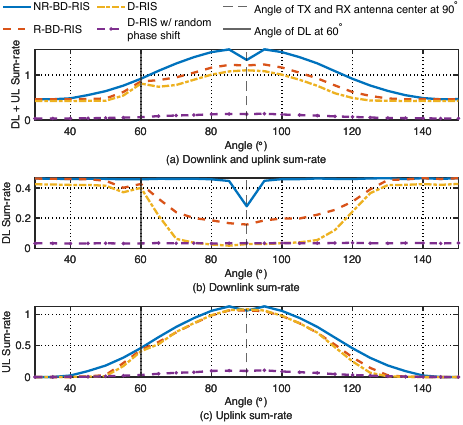}
    \caption{{Comparison of DL and UL sum-rates for NR-BD-RIS, R-BD-RIS, and D-RIS as a function of the UL device angle. The Tx and Rx antennas are positioned at $88^\circ$ and $92^\circ$, respectively, while the DL device is fixed at $60^\circ$. The RIS is configured as a UPA with $M=16$ elements ($4 \times 4$).}}
    \label{fig:angle_rate}
\end{figure}

\subsection{Single-user (SU) Case}

\subsubsection{DL and UL Sum-rate versus the location of UL User}
To observe the sum-rate performance of NR-BD-RIS-assisted FD LEO satellite communication system, we first evaluate the DL and UL sum-rate versus the angle of the UL device. The RIS array is set to a UPA. In addition, the DL device is fixed at $60^\circ$. The UL device is varied from $30^\circ$ to $150^\circ$, this is decided by the height of the satellite and the radius of the Earth. Fig. \ref{fig:angle_rate} presents the results. The round shape is due to that the Earth surface is not flat, thus the pathloss at the edges (\ie $30^\circ$ and $150^\circ$) is large. The results show that NR-BD-RIS performs better than both R-BD-RIS and D-RIS when the UL device is not aligned with the DL device, The reason of the performance gap is that the NR-BD-RIS breaks the channel reciprocity \cite{liu2024non}, thus multiple impinging and reflected directions can be supported. When UL device is aligned with DL device, \ie $60^\circ$, there are just two impinging and reflected directions, thus other two types of RIS can tackle with. In addition, when the angle of UL device is aligned with the Tx antenna at the LEO satellite, there is a deterioration of all the three RISs. This is because that it is difficult for the receive antenna at the LEO satellite to distinguish the signals, and the interference is severe.

\subsubsection{The Impact of the Separation of Tx and Rx}
\begin{figure}
    \centering
    \includegraphics[width=1\linewidth]{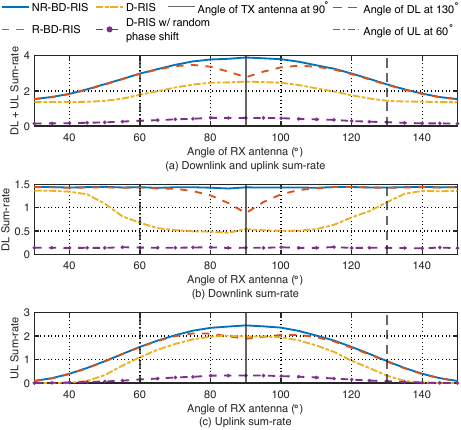}
    \caption{{Comparison of DL and UL sum-rates for NR-BD-RIS, R-BD-RIS, and D-RIS as a function of the UL device angle. The RIS is configured as a UPA with $M=16$ elements ($4 \times 4$). The Tx antenna is fixed at $90^\circ$.}
    \label{fig:angle_rate_ura_4angle}}
\end{figure}
The spacing between the Tx and Rx antennas on the LEO satellite can reduce SI in the propagation domain \cite{everett2016softnull}. To determine how large this spacing can be, we study the DL and UL sum-rates as a function of the angular separation between the Tx and Rx antennas, denoted as $2\Delta_\mathsf{deg}$. Specifically, the Tx antenna is fixed at $90^\circ$, while the Rx antenna is varied from $30^\circ$ to $150^\circ$. The results show that when the separation is within $15^\circ$, the NR-BD-RIS consistently outperforms the R-BD-RIS and D-RIS. This is because the high channel correlation of the DL and UL channels between the RIS and Tx/Rx antennas allows the NR-BD-RIS to effectively maximize the DL and UL sum-rates simultaneously. For separations greater than $15^\circ$, the performance of the NR-BD-RIS is the same as that of the R-BD-RIS, both of which outperform the D-RIS. 
This comes from the interconnected impedance structure of the BD-RIS, which provides more control than the single-connected impedance in the D-RIS and thus leads to better performance. Hence, to limit SI while preserving the gain of NR-BD-RIS, the Tx and Rx antenna separation should stay within $15^\circ$ for this configuration.

\subsubsection{The Impact of Update Interval of NR-BD-RIS}
\begin{figure}
    \centering
    \includegraphics[width=0.85\linewidth]{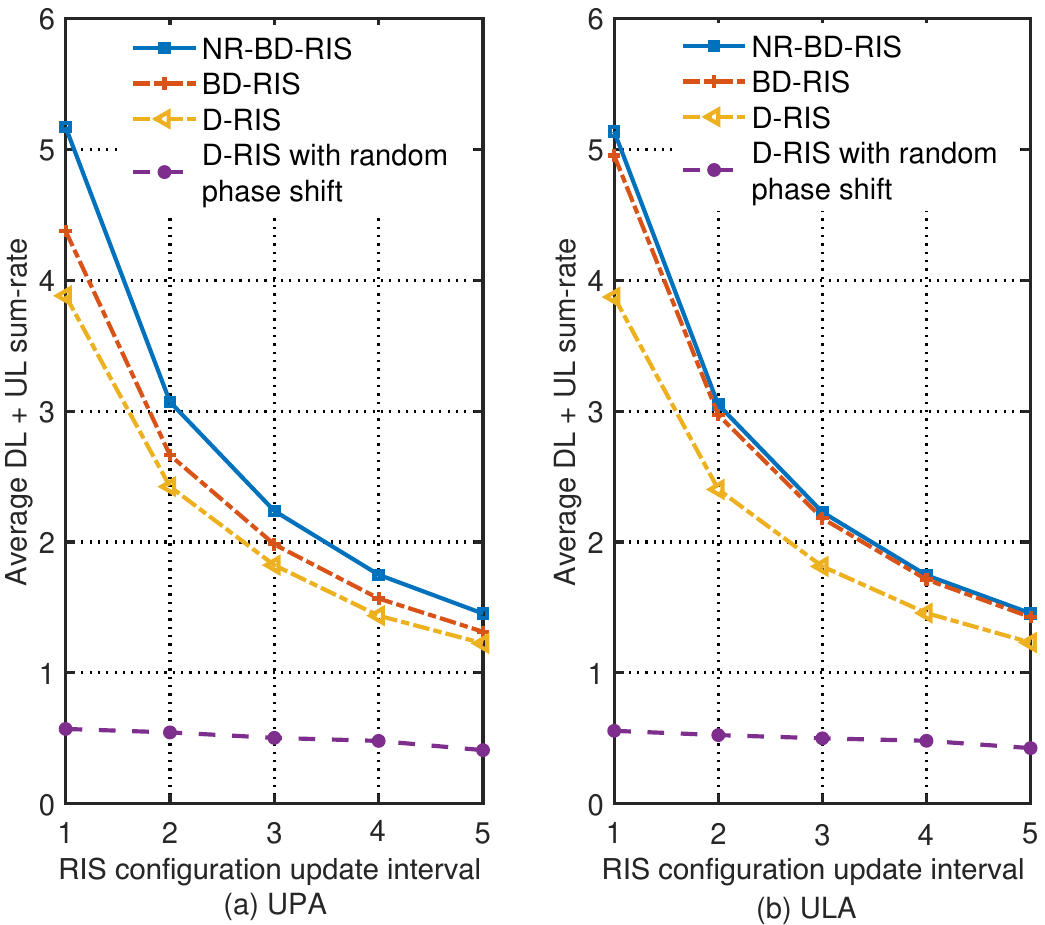}
    \caption{{Average DL and UL sum-rate of NR-BD-RIS, R-BD-RIS, and D-RIS. The locations of the Tx and Rx antenna are fixed at $88^\circ$, and $92^\circ$.  The locations of the DL and UL devices are randomly selected from $[30^\circ, 150^\circ]$. The RIS arrays with $M=32$ RIS elements are set to (a) UPA ($4 \times 8$), (b) ULA ($1 \times 32$).}}
    \label{fig:L}
\end{figure}
\begin{figure}
    \centering
    \includegraphics[width=0.8\linewidth]{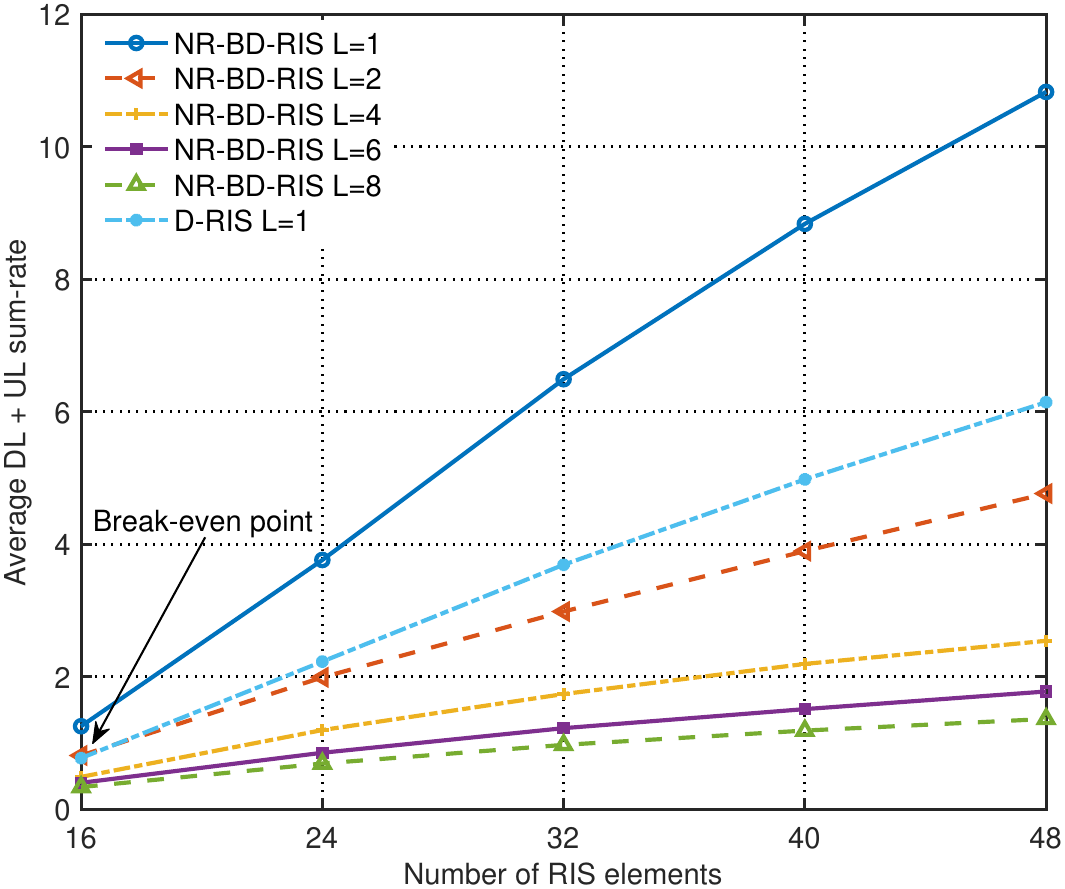}
    \caption{{Average DL and UL sum-rate of the static NR-BD-RIS with update interval $L$ and dynamically updated D-RIS $L=1$. The RIS is set to a UPA with fixed 2 rows. The locations of the Tx and Rx antenna are fixed at $88^\circ$, and $92^\circ$. The locations of the DL and UL devices are located  at $60^\circ$ and $110^\circ$, respectively.}}
    \label{fig:L_ris}
\end{figure}
To alleviate the burden of real-time updates to the scattering matrix, we evaluate the impact of the RIS update interval on the average DL and UL sum-rate over $L$ time slots. The update interval denotes the number of consecutive time slots over which the same RIS configuration is reused. An update interval of $1$ implies that the RIS is reconfigured at every time slot, whereas an update interval of $L$ indicates that the RIS is optimized once and applied to $L$ successive time slots. Specifically, we optimize the scattering matrix by solving $\mathcal{P} 1$, assuming that the locations of all users in the $L$ time slots are known during optimization. The average DL and UL sum-rate per time slot is then calculated. The locations of the DL and UL devices are randomly selected within the range $[30^\circ, 150^\circ]$. 

The results in Fig. \ref{fig:L} show that the NR-BD-RIS always achieves higher performance than the R-BD-RIS and D-RIS, regardless of the RIS update interval. Notably, the performance gap remains even when the RIS is updated less frequently, \ie $L=5$. This highlights the NR-BD-RIS's ability to effectively support devices at varying locations over multiple time slots. This is because of its internal non-reciprocal impedance network. This capability significantly reduces the need for real-time reconfiguration, enhancing its practicality for implementation. In contrast, other two types of RIS show limited flexibility in serving devices over multiple time slots. Thus, they require a shorter update interval to achieve performance comparable to the NR-BD-RIS.

{In addition, we compare the performance of the static NR-BD-RIS with update interval $L$ and the D-RIS that is dynamically updated at every time slot ($L=1$) in Fig. \ref{fig:L_ris}. The break-even point is at $M=16$ RIS elements. When the RIS has more than $16$ elements, the dynamic D-RIS outperforms the static NR-BD-RIS. This is because the D-RIS can fully optimize its configuration for each time slot, while the NR-BD-RIS has a fixed configuration over multiple time slots. Therefore, for scenarios with a limited number of RIS elements, the static NR-BD-RIS is a viable option to reduce the complexity of real-time updates.}

\subsubsection{Sum-rate Region}
\label{sec:sum_region}
\begin{figure}
    \centering
    \includegraphics[width=0.8\linewidth]{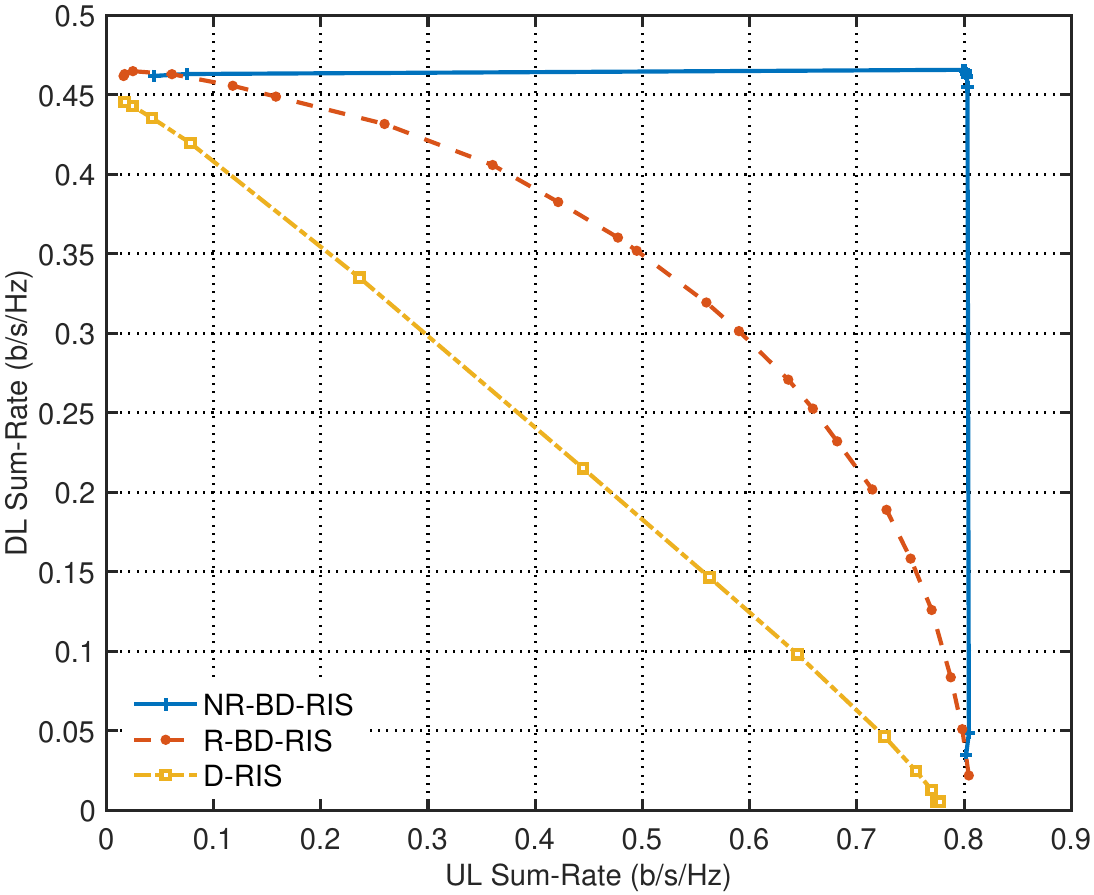}
    \caption{The sum-rate regions for NR-BD-RIS, R-BD-RIS, and D-RIS with $L=1$. The locations of the Tx and Rx antenna are fixed at $88^\circ$, and $92^\circ$. The locations of the DL and UL devices are located  at $60^\circ$ and $110^\circ$, respectively. The RIS with $M=16$ RIS elements is set to a UPA ($4 \times 4$).}
    \label{fig:region_su}
\end{figure}
By varying the DL and UL priority parameters $\alpha_d$ and $\alpha_u$, we can obtain the sum-rate region for the NR-BD-RIS, R-BD-RIS, and D-RIS in Fig. \ref{fig:region_su}. The NR-BD-RIS yields a much larger sum-rate region than the R-BD-RIS and D-RIS, especially when both DL and UL transmissions are considered. This is because the NR-BD-RIS can break channel reciprocity, allowing it to optimally and simultaneously maximize both DL and UL rates. Since the NR-BD-RIS can better tackle with the loop interference and SI, it can achieve true FD operation without a trade-off between DL and UL. When the priority is solely on DL or UL, NR-BD-RIS has no gain over R-BD-RIS, as both can effectively maximize the rate in a single direction. However, both outperform the D-RIS due to the interconnected impedance network of the BD-RIS.

\subsubsection{The Impact of the Number of RIS Elements}
\begin{figure}
    \centering
    \includegraphics[width=0.8\linewidth]{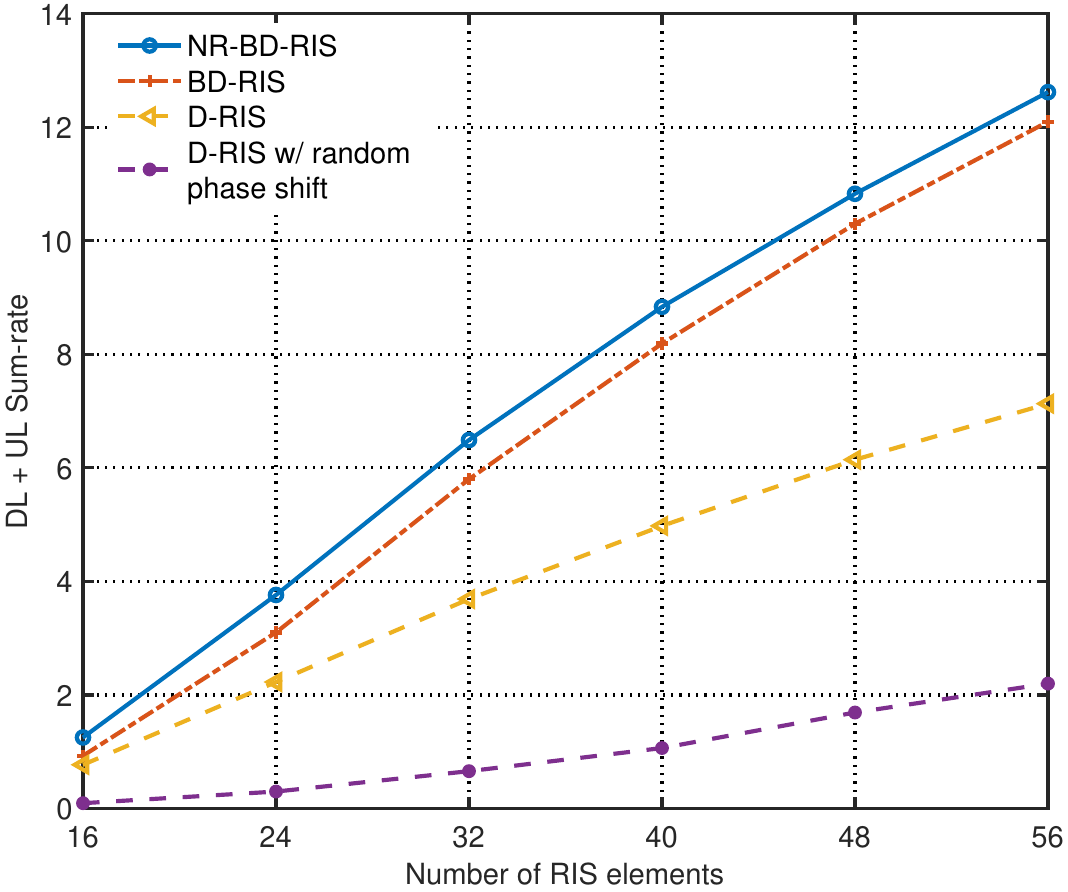}
    \caption{{The DL and UL sum-rate of NR-BD-RIS, R-BD-RIS, and D-RIS versus the number of RIS element with $L=1$. The RIS is set to a UPA with fixed 2 rows. The locations of the Tx and Rx antenna are fixed at $88^\circ$, and $92^\circ$. The locations of the DL and UL devices are located  at $60^\circ$ and $110^\circ$, respectively.}}
    \label{fig:ris}
\end{figure}

To assess how the number of RIS elements affects the DL and UL sum-rate, we analyze the performance as the number of RIS elements increases. As illustrated in \fig{fig:ris}, the NR-BD-RIS consistently outperforms both the R-BD-RIS and D-RIS. This advantage arises from the NR-BD-RIS's ability to break channel reciprocity, allowing it to handle multiple impinging and reflected directions for both DL and UL devices. In contrast, other two types of RIS are limited by channel reciprocity, which restricts their performance capabilities.


\subsubsection{The Impact of the Number of Group Size}
\begin{figure}
    \centering
    \includegraphics[width=0.8\linewidth]{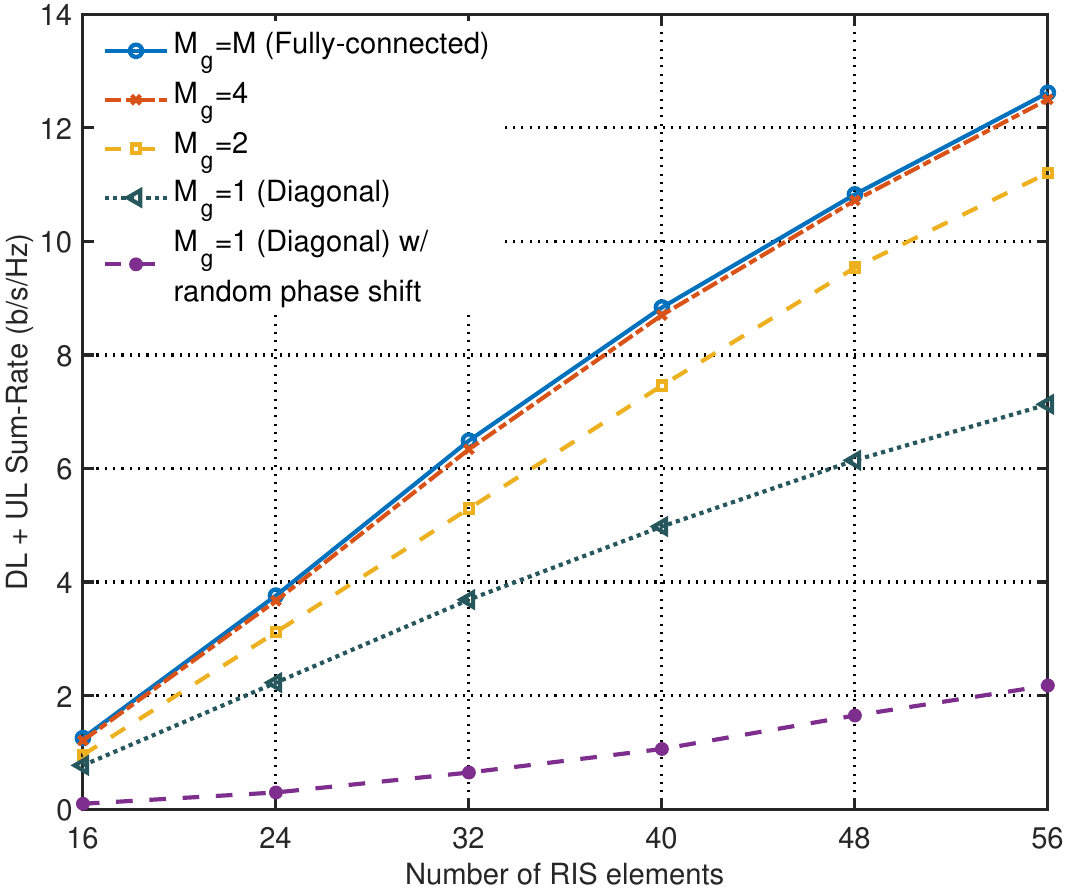}
    \caption{{The DL and UL sum-rate of NR-BD-RIS versus the number of group size with $L=1$. The RIS is set to a UPA with fixed 2 rows. The locations of the Tx and Rx antenna are fixed at $88^\circ$, and $92^\circ$. The locations of the DL and UL devices are located  at $60^\circ$ and $110^\circ$, respectively.}}
    \label{fig:group_nr}
\end{figure}

\begin{figure}
    \centering
    \includegraphics[width=0.85\linewidth]{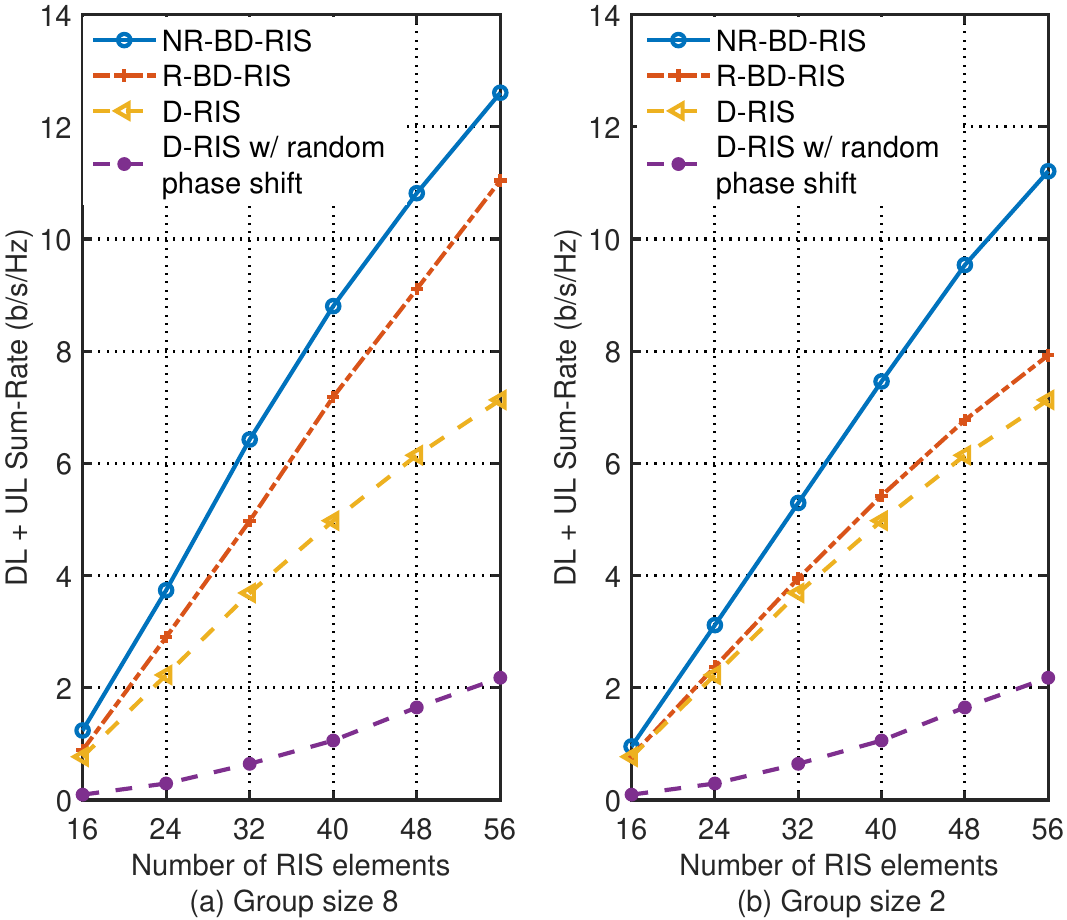}
    \caption{{The DL and UL sum-rate of NR-BD-RIS, R-BD-RIS, and D-RIS with $L=1$ when (a) group size $M_g = 8$ and (b) group size $M_g = 2$. The RIS is set to a UPA with fixed 2 rows. The locations of the Tx and Rx antenna are fixed at $88^\circ$, and $92^\circ$. The locations of the DL and UL devices are located  at $60^\circ$ and $110^\circ$, respectively.}}
    \label{fig:group_8_2}
\end{figure}
We evaluate the impact of group size on the DL and UL sum-rate for the NR-BD-RIS, R-BD-RIS, and D-RIS. The RIS is divided into $G$ groups, each containing $M_g$ elements, such that $M = G M_g$. The group size $M_g$ is varied with a fixed total number of RIS elements $M=56$. When $M_g=1$, it corresponds to the single-connected case, while $M_g=M$ represents the fully-connected case. The results are shown in \fig{fig:group_nr} and \fig{fig:group_8_2}. The DL and UL sum-rate increases with the group size for all three RIS types. This is because a larger group size provides more flexibility in controlling the scattering matrix, thereby enhancing performance. 
Importantly, the NR-BD-RIS consistently achieves better performance than both the R-BD-RIS and D-RIS for all group sizes. This superiority stems from the NR-BD-RIS's capability to break channel reciprocity, enabling it to efficiently handle multiple impinging and reflected directions for both DL and UL devices.

\subsubsection{The Impact of the Power of SI level.}
\begin{figure}
    \centering
    \includegraphics[width=0.8\linewidth]{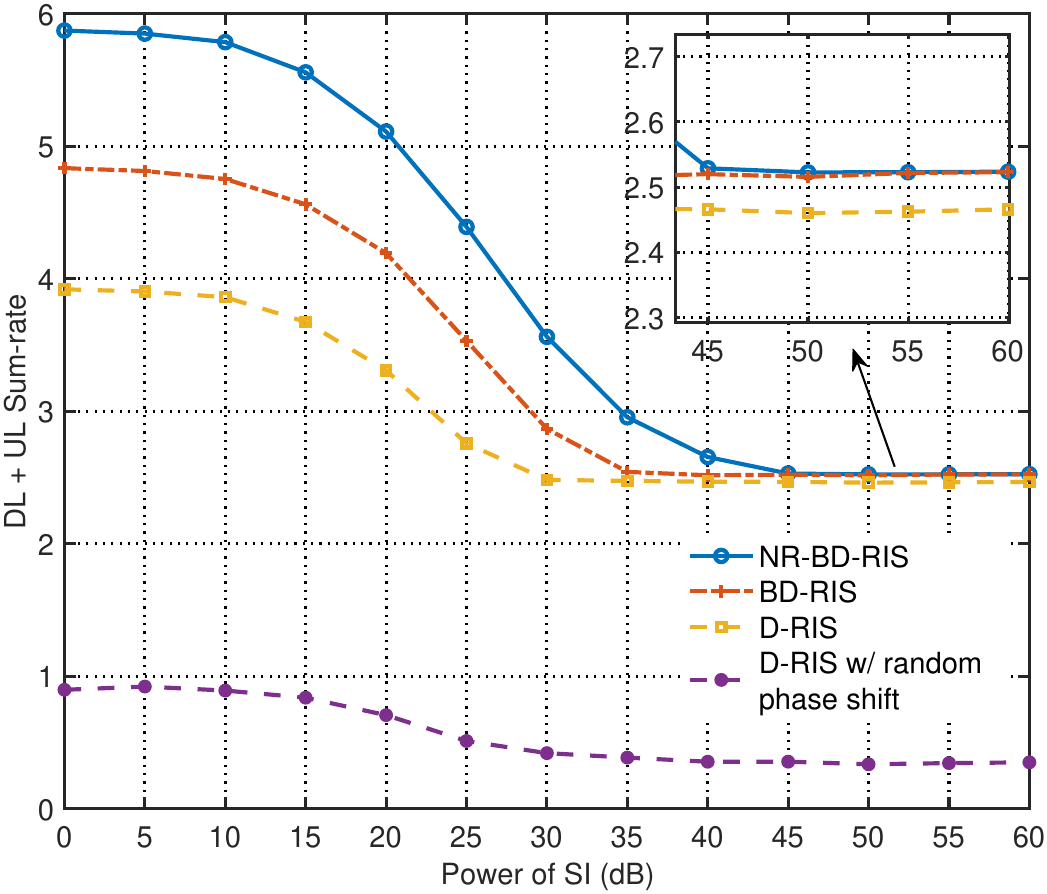}
    \caption{{The DL and UL sum-rate of NR-BD-RIS, R-BD-RIS, and D-RIS versus the number of DL and UL users with $L=1$. The locations of the DL and UL devices are located  at $60^\circ$ and $110^\circ$, respectively. The RIS arrays with $M=16$ RIS elements are set to a ULA ($1 \times 16$).}}
    \label{fig:si}
\end{figure}
We analyze the impact of SI power on the DL and UL sum-rates for the NR-BD-RIS, R-BD-RIS, and D-RIS, with the DL device and UL device located at $60^\circ$ and $110^\circ$, respectively. The Tx/Rx antennas are configured with $N = 1$, and the RIS consists of $M = 16$ elements.
In \fig{fig:si}, it is observed that the sum-rates decline as SI power increases due to the severe interference in the UL SINR \eqref{eq:sinr_ul}. The NR-BD-RIS achieves the highest sum-rates; however, its advantage diminishes as SI power rises. At high SI levels, \eg $45$ dB, the UL rate becomes small, therefore, the DL and UL sum-rate is dominated by the DL rate. This explains why the curves become flat, which reflect the DL rates. As discussed in \ref{sec:sum_region}, when focusing on one-directional transmission, NR-BD-RIS has benefit. Thus, the performances of NR-BD-RIS and R-BD-RIS are the same at high SI levels. The R-BD-RIS outperforms the D-RIS due to the interconnections inside the impedance network, enabling control over both amplitude and phase.

\subsubsection{The Impact of the Number of DL/UL users}
\begin{figure}
    \centering
    \includegraphics[width=0.8\linewidth]{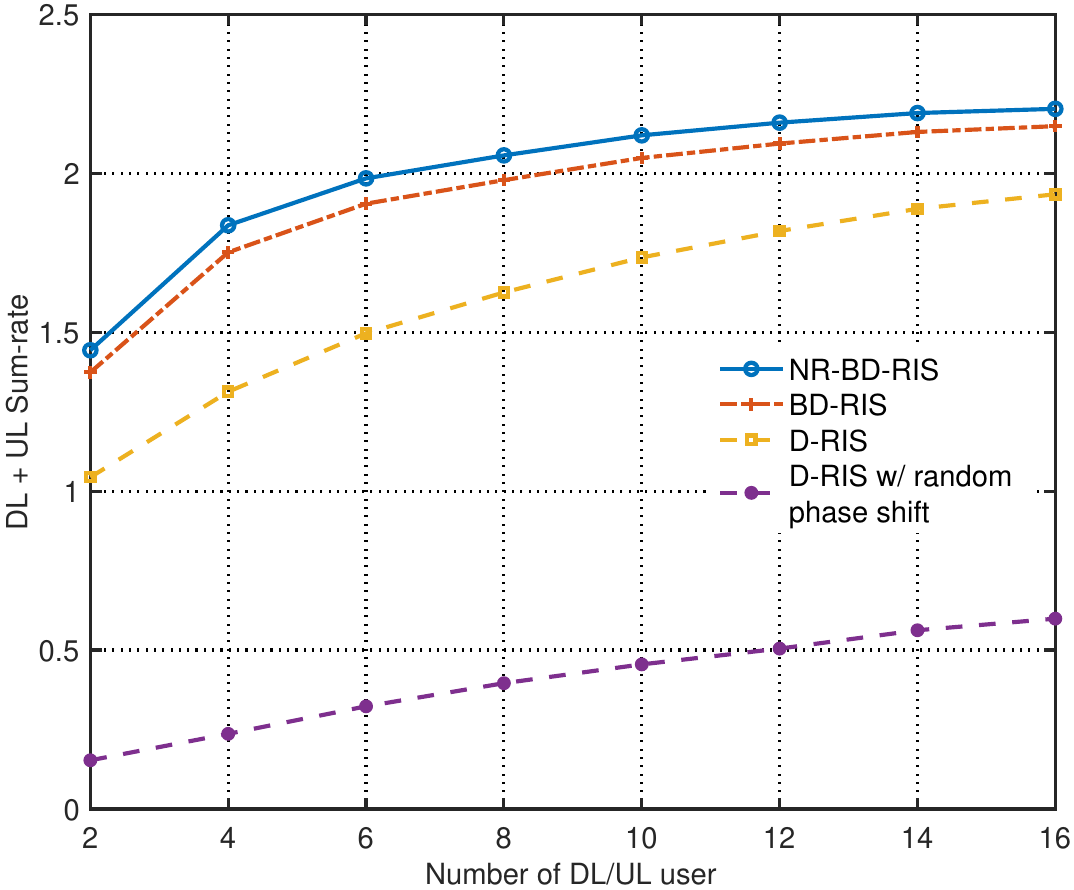}
    \caption{{The DL and UL sum-rate of NR-BD-RIS, R-BD-RIS, and D-RIS versus the number of DL and UL users with $L=1$. The locations of the DL and UL devices are randomly selected from $[30^\circ, 150^\circ]$. The RIS arrays with $M=16$ RIS elements are set to a ULA ($1 \times 16$).}}
    \label{fig:sr_user}
\end{figure}
We analyze the impact of the number of DL and UL users on the DL and UL sum-rate for the NR-BD-RIS, R-BD-RIS, and D-RIS. The Tx/Rx antennas are configured with $N = 1$, and the RIS consists of $M = 16$ elements. The DL and UL devices are randomly positioned within the range $[30^\circ, 150^\circ]$. In \fig{fig:sr_user}, the NR-BD-RIS consistently achieves higher performance compared to the R-BD-RIS and D-RIS for all user counts. While the DL and UL sum-rate increases with the number of users for all three RIS types, the rate of increase diminishes as the number of users grows due to heightened interference among users. The NR-BD-RIS's capability to break channel reciprocity enables it to efficiently handle the added complexity of serving multiple users, thereby sustaining its performance advantage over the other RIS types.

\subsection{Multiple-user (MU) Case}
\subsubsection{The Impact of Update Interval of NR-BD-RIS}
\begin{figure}
    \centering
    \includegraphics[width=0.85\linewidth]{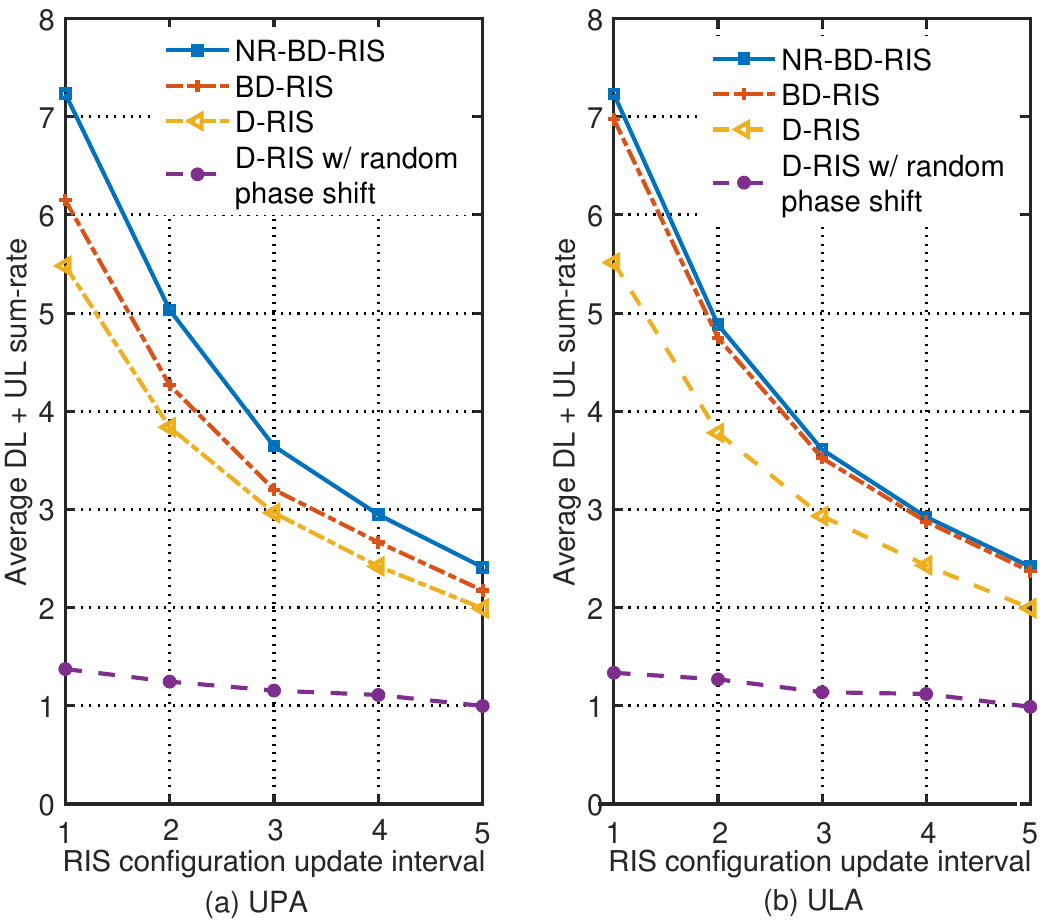}
    \caption{{Average DL and UL sum-rate of NR-BD-RIS, R-BD-RIS, and D-RIS versus the RIS update interval $L$. The Tx and Rx antenna array are fixed at $88^\circ$ and $92^\circ$, respectively. Each time slot serves $G_{d,p} = 2$ DL devices and $G_{u,q} = 2$ UL devices, with their locations randomly selected within $[30^\circ, 150^\circ]$. Both the Tx and Rx arrays have $N=2$ antennas. The RIS is configured as (a) UPA with $M=32$ elements ($4 \times 8$), (b) ULA with $M=32$ elements ($1 \times 32$).}}
    \label{fig:L_mu}
\end{figure}
We analyze the impact of the RIS update interval on the average DL and UL sum-rate over $L$ time slots in a MU scenario. Each time slot accommodates $G_{d,p} = 2$ DL devices and $G_{u,q} = 2$ UL devices, with their positions randomly distributed within $[30^\circ, 150^\circ]$. The Tx and Rx antennas are configured with $N = 2$, and the RIS consists of $M = 32$ elements. The results, presented in Fig. \ref{fig:L_mu}, indicate that the NR-BD-RIS consistently outperforms the R-BD-RIS and D-RIS, regardless of the RIS update interval.
The performance gap remains even when the RIS is updated less frequently, \ie $L=5$. This shows the NR-BD-RIS's ability to effectively support multiple devices at varying locations over multiple time slots due to its non-reciprocal impedance network. This capability significantly reduces the need for real-time reconfiguration, enhancing its practicality for implementation. 

\subsubsection{Sum-rate Region}
\begin{figure}
    \centering
    \includegraphics[width=1\linewidth]{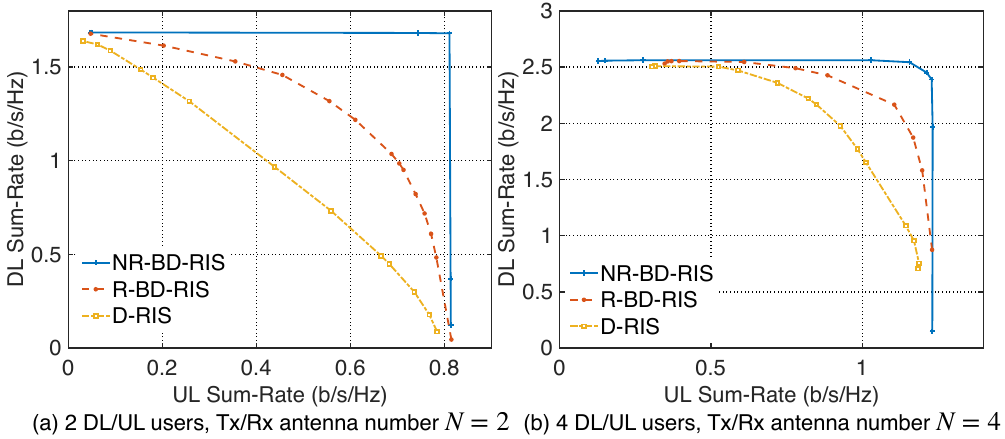}
    \caption{Sum-rate regions for NR-BD-RIS, R-BD-RIS, and D-RIS with $L=1$. The Tx and Rx antennas are fixed at $88^\circ$ and $92^\circ$, respectively. The RIS has $M=16$ elements configured as a UPA ($4 \times 4$). (a) Two DL devices at $80^\circ$ and $140^\circ$, and two UL devices at $40^\circ$ and $110^\circ$. (b) Four fixed DL and UL devices.}
    \label{fig:region_mu}
\end{figure}

The sum-rate regions for the NR-BD-RIS, R-BD-RIS, and D-RIS in a MU scenario are shown in Fig. \ref{fig:region_mu}. In Fig. \ref{fig:region_mu}(a), two DL devices are located at $80^\circ$ and $140^\circ$, while two UL devices are at $40^\circ$ and $110^\circ$. In Fig. \ref{fig:region_mu}(b), four fixed DL and UL devices are considered. The sum-rate region of the NR-BD-RIS is still larger than that of the R-BD-RIS and D-RIS in MU cases, especially when both DL and UL transmissions are considered. This is because the NR-BD-RIS can break channel reciprocity, allowing it to simultaneously maximize both DL and UL rates. When one direction transmission is considered, NR-BD-RIS has no gain over R-BD-RIS, as both can effectively maximize the rate in a single direction. However, both outperform the D-RIS due to the interconnected impedance network of the BD-RIS.

\section{Conclusion}
\label{sec:conclusion}
This work has investigated the FD LEO satellite communication system enabled by NR-BD-RIS to achieve simultaneous DL and UL transmission. To serve a broader range of ground devices, we have considered a time-sharing scheduling framework. We have formulated the weighted DL and UL sum-rate maximization problem over time. To solve the non-convex problem, we have designed an iterative algorithm using BCD and PDD methods. The numerical results have shown that the NR-BD-RIS outperforms the conventional R-BD-RIS and D-RIS, especially when the RIS is updated less frequently. In addition, the gain of NR-BD-RIS keeps even when the Tx and Rx antennas are not strictly aligned. In MU scenarios, the NR-BD-RIS demonstrates gains and performance in supporting multiple users with varying locations. We attribute these advantages to the NR-BD-RIS's ability to break channel reciprocity, enabling it to effectively control multiple impinging and reflected directions for both DL and UL devices.
{Future research directions include rigorously evaluating the energy efficiency of NR-BD-RIS by explicitly accounting for the additional power consumption introduced by non-reciprocal components, and investigating NR-BD-RIS design under imperfect CSI conditions to enhance robustness in practical scenarios.}
This work paves the way for future studies on NR-BD-RIS-enabled satellite communications. 
\bibliographystyle{IEEEtran_url}
\bibliography{IEEEabrv,references}

\begin{IEEEbiography}
[{\includegraphics[width=1in,height=1.25in,clip,keepaspectratio]{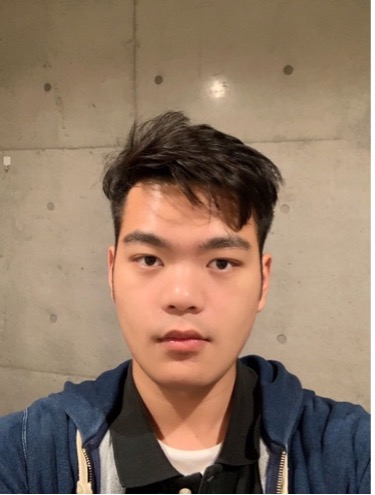}}]%
{Ziang Liu} is currently a postdoctoral researcher at Imperial College London, UK. He received his B.Eng. degree in Communication Engineering from the University of Electronic Science and Technology of China (UESTC), Chengdu, China, in 2018. He earned his M.Eng. degree in Information and Communications Engineering from the Tokyo Institute of Technology, Tokyo, Japan, in 2020, and obtained his Ph.D. in Communications and Signal Processing at the Department of Electrical and Electronic Engineering, Imperial College London, UK. He was a visiting Ph.D. student at Cornell University. He is the recipient of the 2024 IEEE Transportation Electronics Fellowship Award and the second place in the 2025 IEEE IoT PhD Thesis Competition. His research interests include integrated sensing and communications, modulo sampling-based ADC, and optimization.
\end{IEEEbiography}

\begin{IEEEbiography}
[{\includegraphics[width=1in,height=1.25in,clip,keepaspectratio]{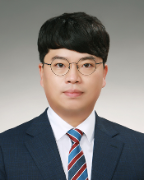}}]%
{Wonjae Shin} (Senior Member, IEEE) received the B.S. and M.S. degrees from the Korea Advanced Institute of Science and Technology (KAIST) in 2005 and 2007, respectively, and the Ph.D. degree from the Department of Electrical and Computer Engineering, Seoul National University (SNU), South Korea, in 2017. He has been a Postdoctoral Research Fellow and a Visiting Scholar at Princeton University, Princeton, NJ, USA, from 2016 to 2018. From 2007 to 2014, he was a Member of the Technical Staff at the Samsung Advanced Institute of Technology and Samsung Electronics Co., Ltd., South Korea, where he contributed to next-generation wireless communication networks, particularly to 3GPP LTE/LTE-Advanced standardization. Since 2023, he has been with the School of Electrical Engineering, Korea University, Seoul, South Korea, where he is currently an Associate Professor. Prior to joining Korea University, he was a Faculty Member at Pusan National University and Ajou University, South Korea. His research interests include designing and analyzing future wireless communication systems, such as interference-limited networks, non-terrestrial networks (NTN), and AI/ML-aided wireless networks. Dr. Shin was the recipient of the Fred W. Ellersick Prize and the Asia-Pacific Outstanding Young Researcher Award from the IEEE Communications Society in 2020, the Haedong Young Engineer Award in 2025, the JCN Best Paper Award in 2025, the ICTC Best Workshop Paper Award in 2022, the J-KICS Best Paper Award in 2021, the Best Ph.D. Dissertation Award from SNU in 2017, the Gold Prize from the IEEE Student Paper Contest (Seoul Section) in 2014, and the Ministry of Science and ICT of Korea Award in IDIS-Electronic News ICT Paper Contest in 2017. He serves as an Associate Editor for the IEEE INTERNET OF THINGS JOURNAL and the IEEE TRANSACTIONS ON VEHICULAR TECHNOLOGY.
\end{IEEEbiography}

\begin{IEEEbiography}
[{\includegraphics[width=1in,height=1.25in,clip,keepaspectratio]{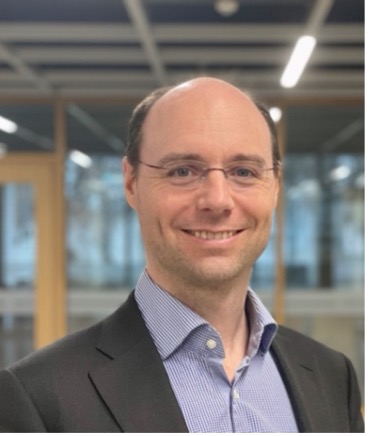}}]%
{Bruno Clerckx} (Fellow, IEEE) received the M.Sc. and Ph.D. degrees in electrical engineering from the Universit\'e catholique de Louvain, Belgium, and the Doctor of Science (D.Sc.) degree from the Imperial College London, London, U.K. Before joining Imperial College London in 2011, he was with Samsung Electronics, Suwon-si, South Korea, where he actively contributed to 4G (3GPP LTE/LTE-A and IEEE 802.16m). He was also the Chief Technology Officer (CTO) of Silicon Austria Laboratories (SAL) from 2023 to 2024. He is currently a (Full) Professor, the Head of the Wireless Communications and Signal Processing (WCSP) Laboratory, and the Head of the Communications and Signal Processing Group, Department of Electrical and Electronic Engineering, Imperial College London. He has authored two books on MIMO Wireless Communications and MIMO Wireless Networks, 300 peer-reviewed international research articles, and 150 standards contributions, and he is the inventor of 80 issued or pending patents, among which several have been adopted in the specifications of 4G standards and are used by billions of devices worldwide. His research spans the general area of wireless communications and signal processing for wireless networks. He is a fellow of the IET. He received the prestigious Blondel Medal from France for exceptional work contributing to the progress of Science and Electrical and Electronic Industries, the 2021 Adolphe Wetrems Prize in mathematical and physical sciences from the Royal Academy of Belgium, multiple awards from Samsung, the IEEE Best Student Paper Award, and European Association for Signal Processing (EURASIP) Best Paper Award 2022. He is an IEEE Communications Society Distinguished Lecturer.
\end{IEEEbiography}

\end{document}